\documentclass[twocolumn]{aastex62}
\accepted{to AAS journals on 01/19/2021}
\received{08/26/2020}
\usepackage{graphicx}
\usepackage{natbib}
\usepackage{latexsym}
\usepackage{amssymb}
\usepackage{longtable}
\usepackage{amsmath}
\usepackage{url}
\def\msun{\ifmmode {\rm\,M_\odot}\else ${\rm\,M_\odot}$\fi}
\def\Msun{\ifmmode {\rm\,\it{M_\odot}}\else ${\rm\,M_\odot}$\fi}
\def\lsun{\ifmmode {\rm\,L_\odot}\else ${\rm\,L_\odot}$\fi}
\def\Lsun{\ifmmode {\rm\,\it{L_\odot}}\else ${\rm\,L_\odot}$\fi}
\def\rsun{\ifmmode {\rm\,R_\odot}\else ${\rm\,R_\odot}$\fi}
\def\Rsun{\ifmmode {\rm\,\it{R_\odot}}\else ${\rm\,R_\odot}$\fi}
\def\Tsun{\ifmmode {\rm\,T_\odot}\else ${\rm\,T_\odot}$\fi}
\def\arcsec{\ifmmode {^{\prime\prime}}\else $^{\prime\prime}$\fi}
\def\asec{\ifmmode {^{\prime\prime}}\else $^{\prime\prime}$\fi}
\def\arcmin{\ifmmode {^{\prime}}\else $^{\prime}$\fi}
\def\amin{\ifmmode {^{\prime}}\else $^{\prime}$\fi}
\def\simlt{\mathrel{\spose{\lower 3pt\hbox{$\mathchar"218$}}
     \raise 2.0pt\hbox{$\mathchar"13C$}}}
\def\simgt{\mathrel{\spose{\lower 3pt\hbox{$\mathchar"218$}}
\     \raise 2.0pt\hbox{$\mathchar"13E$}}}

\def\halpha{H$\alpha$}
\def\hbeta{H$\beta$}

\begin{document}

\author{P. Wilson Cauley}
\affiliation{Laboratory for Atmospheric and Space Physics, University of Colorado Boulder, Boulder, CO 80303}

\author{Ji Wang}
\affiliation{Department of Astronomy, Ohio State University, Columbus, OH 43210}

\author{Evgenya L. Shkolnik}
\affiliation{Arizona State University, School of Earth and Space Exploration, Tempe, AZ 85287}

\author{Ilya Ilyin}
\affiliation{Leibniz-Institute for Astrophysics Potsdam (AIP), An der Sternwarte 16, 14482, Potsdam, Germany}

\author{Klaus G. Strassmeier}
\affiliation{Leibniz-Institute for Astrophysics Potsdam (AIP), An der Sternwarte 16, 14482, Potsdam, Germany}

\author{Seth Redfield}
\affiliation{Wesleyan University, Astronomy Department, Van Vleck Observatory, Middletown, CT 06459}

\author{Adam G. Jensen}
\affiliation{Department of Physics and Astronomy, University of Nebraska at Kearney, Kearney, NE 68849}

\correspondingauthor{P. Wilson Cauley}
\email{pwcauley@gmail.com}

\title{Time-resolved rotational velocities in the upper atmosphere of
WASP-33 b\footnote{Based on data acquired with PEPSI
using the Large Binocular Telescope (LBT). The LBT is an international
collaboration among institutions in the United States, Italy, and Germany. LBT
Corporation partners are the University of Arizona on behalf of the Arizona
university system; Istituto Nazionale di Astrofisica, Italy; LBT
Beteiligungsgesellschaft, Germany, representing the Max-Planck Society, the
Leibniz-Institute for Astrophysics Potsdam (AIP), and Heidelberg University;
the Ohio State University; and the Research Corporation, on behalf of the
University of Notre Dame, University of Minnesota and University of Virginia.}}

\begin{abstract}

While steady empirical progress has been made in understanding the structure and composition
of hot planet atmospheres, direct measurements of velocity signatures, including
winds, rotation, and jets, have lagged behind. Quantifying atmospheric dynamics
of hot planets is critical to a complete understanding of their atmospheres and
such measurements may even illuminate other planetary properties, such as magnetic
field strengths. In this manuscript we present the first detection of the Balmer lines
H$\alpha$ and H$\beta$ in the atmosphere of the ultra-hot Jupiter WASP-33 b. Using
atmospheric models which include the effects of atmospheric dynamics, we
show that the shape of the average Balmer line transmission spectrum is consistent with rotational 
velocities in the planet's thermosphere of $v_\text{rot} = 10.1^{+0.8}_{-1.0}$ km s$^{-1}$. 
We also measure a low-significance day-to-night side velocity shift of $-4.6^{+3.4}_{-3.4}$ km s$^{-1}$ 
in the transmission spectrum which is naturally explained by a global wind across the planet's 
terminator. In a separate analysis the time-resolved velocity centroids of individual
transmission spectra show unambiguous evidence of rotation, with a best-fit velocity
of $10.0^{+2.4}_{-2.0}$ km s$^{-1}$, consistent with the value of $v_\text{rot}$ derived
from the shape of the average Balmer line transmission spectrum. Our observations 
and analysis confirm the power of high signal-to-noise, time-resolved transmission spectra to 
measure the velocity structures in exoplanet atmospheres. The large rotational 
and wind velocities we measure highlight the need for more detailed 3D global climate 
simulations of the rarefied upper-atmospheres of ultra-hot gas giants.

\end{abstract}

\keywords{}

\section{INTRODUCTION}
\label{sec:intro}

The characterization of hot exoplanet atmospheres has advanced steadily 
over the past decade, with details continuously being revealed about the
diversity of thermal profiles \citep{line16,evans17,nugroho17,nikolov18,gibson20,yan20,baxter20}, 
chemical abundances \citep{line14,brogi19,pino20},
atmospheric evaporation processes \citep{bourrier13,ehrenreich15,lavie17,bourrier18}, 
and the presence of clouds and hazes \citep{kreidberg14,barstow17,moran18,beatty19,libby20,gao20}. 
Recent observations of hot planet atmospheres at high spectral resolution, 
both in transmission and via cross-correlation of thermal emission signatures, 
have accelerated the detection of a wide variety of molecular and atomic species 
\citep[e.g.,][]{casasayas17,jensen18,spake18,allart18,salz18,hoeijmakers18,hoeijmakers19,cauley19,sing19,brogi19,yan19,keles19,vonessen19,turner20,nugroho20,nugroho20a,benyami20,stangret20,cabot20,chen20}. 

The direct observation of velocity signatures has, however, lagged 
behind the otherwise remarkable progress made in understanding the
properties of hot planet atmospheres. Quantifying velocity dynamics
such as rotation, equatorial jets, and hydrodynamic expansion can reveal
crucial information about related planetary properties, such
as magnetic fields \citep{cauley19a}, heat redistribution efficiency and mass loss rates
\citep[][]{showman02,kempton12,spiegel13},
and provide important feedback to global climate simulations (GCMs) 
\citep{showman11,rauscher13,carone20}. Thermal phase curves and low-resolution
spectra, while not directly measuring the Doppler shifts of atmospheric gas,
have been used to infer the presence of heat redistribution by
winds and jets \citep[e.g.,][]{knutson12,kataria15,wong16,rogers17,wong20,vonessen20}.
Wind speeds have also been inferred for brown dwarfs using IR
and radio variability \citep{apai17,allers20}.
In addition to the relative dearth of observed velocity signatures, there is 
also a critical lack of GCMs which take into account the
most rarefied bound atmospheric layers at pressures $p < 10^{-3}$ 
bar, i.e. the thermosphere, that are typically sampled with transmission 
spectroscopy of atomic ions.

The first measurement of rotational broadening in an exoplanet
atmosphere was performed by \citet{snellen14} who constrained
the equatorial rotation velocity of the young planet $\beta$ Pic b
to $v_\text{rot} = 25.0 \pm 3.0$ km s$^{-1}$ using emission features
of CO and H$_2$O. Such large rotation velocities are generally only 
possible for young inflated planets which 
have yet to fully contract \citep{baraffe03}. However, the \citet{snellen14}
result demonstrated the feasibility of using emission or absorption
profiles to quantify rotational velocities in exoplanet atmospheres. 
Not long after, studies by \citet{louden15} and \citet{brogi16} were
able to constrain the rotational velocities in the atmosphere
of HD 189733 b and found them to be consistent with the tidally-locked
value of $\approx 2.7$ km s$^{-1}$. 

Despite the relative paucity of clear velocity measurements
in hot planet atmospheres, these signatures are beginning to be
teased apart by detailed examination of transmission spectra
and the application of more sophisticated simulations. 
\citet{flowers19} used a suite of 3D GCMs coupled with a 1D
radiative transfer code to constrain the day-to-night side wind
speed and equatorial rotational velocity of HD 189733 b \citep{louden15,wyttenbach15,salz18}. 
\citet{wyttenbach20} included rotational broadening in their detailed analysis of the
Balmer line transmission spectra for KELT-9 b, showing that the 
profile shapes are consistent with the tidally locked equatorial
rotational velocity. Although the \ion{He}{1} 10830 \AA\ triplet
may trace the unbound exosphere, \citet{allart18} and \citet{allart19} reported
blue-shifted absorption in this line around HAT-P-11 b and WASP-107 b, respectively, 
potentially indicative of day-to-night side winds in the thermosphere. 
\citet{ehrenreich20} demonstrated an unprecedented level of precision in measuring the velocity
centroids of individual absorption profiles for the hot Jupiter
WASP-76 b and found that the highly blue-shifted absorption on the
trailing limb of the planet can be explained by a combination of
winds and rotation. Finally, the most comprehensive application
of velocity flows in modeling a transmission spectrum was recently
published by \citet{seidel20}, who found large expansion velocities
are necessary to explain the shape of the \ion{Na}{1} D absorption line
in HD 189733 b's atmosphere.

In the present manuscript we focus on transmission spectroscopy
of the Balmer lines \halpha\ and \hbeta\ in the atmosphere of 
the ultra-hot Jupiter (UHJ) WASP-33 b \citep{collier10}. WASP-33 b
has an equilibrium temperature of $T_\text{eq} \approx 2750$ K and 
is a known pulsator with pulsation periods of $\approx 1$ hour \citep{herrero11,vonessen14}.
It has recently joined the growing list of hot Jupiters and UHJs with 
atomic detections of their atmospheres \citep{yan19,nugroho20}. We present the
first detection of the Balmer lines in WASP-33 b's atmosphere\footnote{During the review
process we became aware of the Balmer line detection by \citet{yan20a} for WASP-33 b
using data collected with HARPS-N and CARMENES.} and
discuss in detail how measurements of the \halpha\ centroids in 
the transmission spectra, especially those during ingress and egress
of the transit, reveal details about rotational velocities in the planet's 
thermosphere.  

\section{Atmospheric velocities from time-series transmission spectra}
\label{sec:vels}

The average transmission spectrum, or equivalent cross correlation profile 
\citep{brogi16,hoeijmakers18}, can provide information on the broadening mechanisms 
responsible for the width of the line profile
\citep[e.g.,][]{louden15,allart18,allart19,cauley19} and can also reveal
the net blue-shift of a day-to-night side wind
\citep[e.g.,][]{snellen10,wyttenbach15,casasayas19,bourrier20}.
However, most broadening mechanisms (thermal, hydrodynamic expansion, rotation, and jets)
are degenerate to some degree and thus it is difficult to disentangle them
using only the shape of the average in-transit absorption. 

\begin{figure*}[htbp]
   \centering
   \includegraphics[scale=.8,clip,trim=10mm 5mm 0mm 5mm,angle=0]{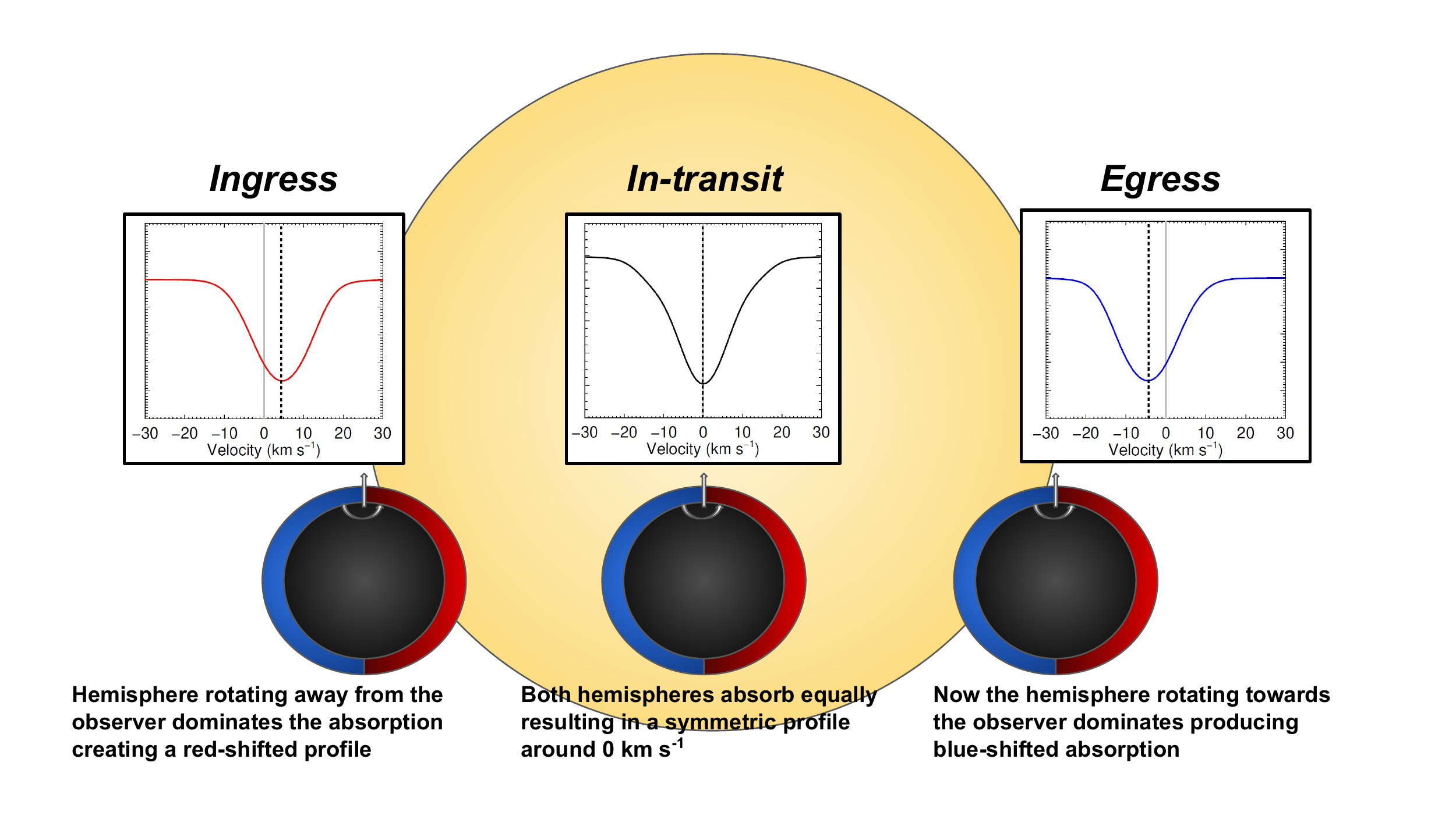}
   \figcaption{Demonstration of how a rotating atmosphere produces red and blue-shifted
   transmission spectra at various transit times. Upon ingress, the planetary hemisphere
   rotating away from the observer dominates the transmission spectrum resulting in
   a red-shifted absorption profile. The mid-transit spectrum exhibits no velocity
   shift but shows maximal broadening from the rotating atmosphere. Finally, the
   hemisphere rotating towards the observer dominates the absorption profile during
   egress, producing a blue-shifted transmission spectrum. Note that contributions
   from the star are ignored in the transmission spectrum examples.
\label{fig:timeres}}

\end{figure*}

\begin{figure}[htbp]
   \centering
   \includegraphics[scale=.4,clip,trim=30mm 30mm 10mm 25mm,angle=0]{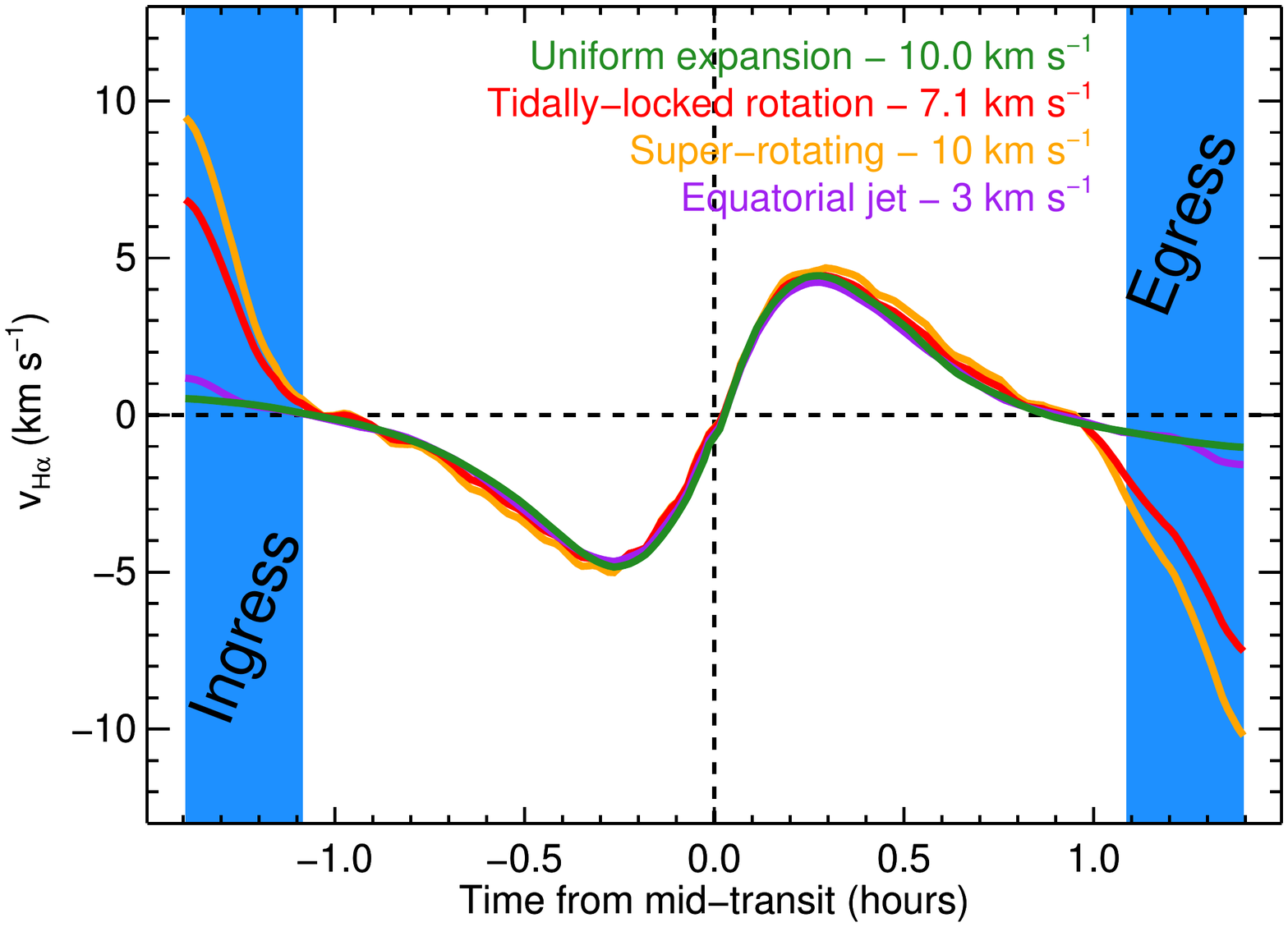}
   \figcaption{Examples of model velocity signatures for WASP-33 b. The uniformly
   expanding atmosphere (green line) shows no velocity shifts during ingress or 
   egress due to the symmetric nature of the velocity profile. The rotating
   atmospheres show red-shifted absorption profiles 
   upon ingress and blue-shifted profiles upon egress. Since the jet is confined to
   equatorial latitudes the velocity shift is weaker than in the rotating atmosphere
   case. Note that the non-zero velocities 
   between ingress and egress are a result of the planet's atmosphere absorbing different portions
   of the stellar \halpha\ line profile, producing slightly asymmetric absorption
   profiles. This occurs independently of the broadening mechanism.
\label{fig:velexamps}}

\end{figure}

Time-series measurements of atmospheric absorption provide a means to
break this degeneracy. Assuming a spherically symmetric atmosphere, 
both thermal broadening and atmospheric expansion produce symmetric
broadening effects on the planet's transmission spectrum regardless of
when during transit the planet is observed. Rotation and, to a lesser degree, jets produce
an asymmetric broadening effect during ingress and egress where, assuming
the planet's spin axis is perpendicular to the plane of its orbit, only
one hemisphere dominates the transmission spectrum. We illustrate this
with a cartoon in \autoref{fig:timeres} \citep[see also Figures 1, 19, or 5 of][respectively]{louden15,cauley17,flowers19}.
We also show the difference between velocity centroids produced by atmospheric 
expansion, rotation, and jets in \autoref{fig:velexamps} for the case of WASP-33 b
(see \autoref{sec:mods} for a description of the models). 
For the rotation case, the velocity centroids of the transmission spectrum 
are blue- or red-shifted depending on which hemisphere produces the absorption. 
This effect was first explored by \citet[][see their Figure 8]{kempton12} who 
showed that such ingress and egress velocity shifts should be detectable for 
hot Jupiter atmospheres.

Until recently time-resolved measurements of transmission spectrum
centroids were difficult to obtain due to the mechanical and thermal instabilities 
inherent to non-climate controlled echelle spectrographs. As the stability
and precision of high-resolution spectrographs has improved, it has become
more feasible to collect high signal-to-noise transmission spectra as a
function of time throughout a transit. The most spectacular example of
this technique was presented by \citet{ehrenreich20} who used the
ESPRESSO spectrograph to measure the velocity centroids of \ion{Fe}{1} 
absorption in the atmosphere of WASP-76 b at a cadence of $\approx 6 - 7$
minutes. Using PEPSI on the LBT, we were able to measure the 
velocity centroids of the \halpha\ absorption in the atmosphere of KELT-9 b
at a $\approx 5$ minute cadence \citep{cauley19}. We note that
\citet{borsa20} also measured the velocity centroids of absorption features
in the atmosphere of WASP-121 b (see their Figure 6) but do not analyze the
time series from the framework of a rotating atmosphere.

We adopt an empirical approach to determining the velocities of our \halpha\ 
transmission spectra. In other words, we do not assume a functional form for
the shape of the lines and instead measure the velocities directly from the
\halpha\ line profiles. The metric we use is the same as that defined in 
\citet{cauley17}:

\begin{equation}\label{eq:vha}
    v_{H\alpha} = \frac{\sum^{v=+50}_{v=-50} v(1-F(v))^2}{\sum^{v=+50}_{v=-50} (1-F(v))^2}
\end{equation}

\noindent where $F(v)$ is the normalized flux in the transmission spectrum
at velocity $v$. Uncertainties for $v_{H\alpha}$ are estimated from the weighting
function. First we sort the weighting function $(1-F(v))^2$ from largest to 
smallest and, beginning with the largest values, we sum the function until 
68\% of the total has been reached. We then calculate the standard deviation 
of the mean of these velocities and take that value as the 1$\sigma$ uncertainty on $v_{H\alpha}$.
Thus broader profiles have larger uncertainties.
The flux-weighted velocity measurement ensures that the
deepest portions of the line profile dominate the centroid determination. 
In general, using $v_{H\alpha}$ results in similar velocity centroids 
compared with, for example, fitting a Gaussian to the line profile. However
when the line is asymmetric, $v_{H\alpha}$ often does a better job of finding the
velocity associated with the region of greatest absorption since the 
profile morphology is not strictly Gaussian. \autoref{eq:vha}
is used to calculate the model velocities in \autoref{fig:velexamps}.
We will return to the $v_{H\alpha}$ measurements of the data in \autoref{sec:obsvel}.

\section{Observations and Data reduction}
\label{sec:obs}

We observed a single transit of WASP-33 b on UT 2019-11-17 using the PEPSI
\citep{strassmeier15} spectrograph on the Large Binocular Telescope (LBT).
The observations began at UT 01:02 and ended at UT 11:19, resulting in
$\approx 4.75$ hours of pre-transit exposures, $\approx 2.0$ hours
of post-transit exposures, and the entire $\approx 2.8$ hour transit. 

PEPSI was used in its $R \approx 50,000$ mode and with cross dispersers (CD)
III (blue arm) and V (red arm) simultaneously. The wavelength coverage was
4750--5430\,\AA\ in the blue arm and 6230--7430\,\AA\ in the red arm. The
spectra were collected with a constant signal-to-noise of 210 pixel$^{-1}$ in the continuum
controlled by a photon counter. The use of the photon counter results in slightly
different exposure times for each spectrum, which ranged from $\approx 5$ minutes
to $\approx 10$ minutes depending on airmass and seeing. A total of 82 spectra were
collected in both the red and blue arms, including 26 in-transit spectra and
56 out-of-transit spectra.

The PEPSI data reduction routines follow standard high-resolution extraction
procedures. Briefly, the individual science images are bias subtracted, flat fielded, and
then optimally extracted. The extracted spectra are normalized using a spline
fit to the continuum and corrected for the Earth's barycentric motion at the time 
of the observation. We also correct for the system velocity in \autoref{tab:pars} 
to place the spectra in the rest frame of the star. More details on PEPSI data 
reduction can be found in \citet{strassmeier18} and \citet{cauley19}.  

\begin{deluxetable*}{lcccc}
\tablecaption{WASP-33 system parameters\label{tab:pars}}
\tablehead{\colhead{Parameter}&\colhead{Symbol}&\colhead{Units}&\colhead{Value}&\colhead{Reference}}
\colnumbers
\tabletypesize{\scriptsize}
\startdata
Stellar mass & $M_\star$ & $M_\Sun$ & $1.561^{+0.045}_{-0.079}$ & 1\\
Stellar radius & $R_\star$ & $R_\Sun$ & $1.509^{+0.016}_{-0.027}$ & 1\\
Stellar surface gravity & log$g$ & cm s$^{-2}$ & $4.3 \pm 0.2$ & 3\\
Effective temperature & $T_\text{eff}$ & K & $7430 \pm 100$ & 1 \\
Metallicity & [Fe/H] & \nodata & $-0.1 \pm 0.2$ & 1\\
Stellar rotational velocity & $v$sin$i$ & km s$^{-1}$ & $86.63^{+0.37}_{-0.32}$ & 2\\
Spin-orbit alignment angle & $\lambda$ & degrees & $-109.29^{+0.20}_{-0.17}$ & This work\\
Orbital period & $P_\text{orb}$ & days & $1.2198669 \pm 0.0000012$ & 3\\
Semi-major axis & $a$ & AU & $0.02390 \pm 0.00063$ & 4 \\
Planetary mass & $M_\text{p}$ & $M_\text{J}$ & $2.16 \pm 0.20$ & 1 \\
Planetary radius & $R_\text{p}$ & $R_\text{J}$ & $1.679^{+0.019}_{-0.030}$ & 6\\
Orbital velocity & $K_\text{p}$ & km s$^{-1}$ & $231 \pm 3$ & 5\\
Mid-transit time & $T_0$ & JD & $2458804.829075^{+0.00051}_{-0.00047}$ & This work\\
Transit duration & $T_{14}$ & hours & $2.7896^{+0.0039}_{-0.0037}$ & This work\\
System velocity & $\gamma$ & km s$^{-1}$ & $4.63 \pm 0.04$ & This work 
\\
\enddata
\tablerefs{$1=$ \citet{lehmann15}; $2=$ \citet{johnson15}; $3=$ \citet{collier10}; $4=$ \citet{chakrabarty19}
$5=$ \citet{yan19}; $6=$ \citet{turner16}}
\end{deluxetable*}

\begin{figure}[htbp]
   \centering
   \includegraphics[scale=.50,clip,trim=50mm 25mm 20mm 45mm,angle=0]{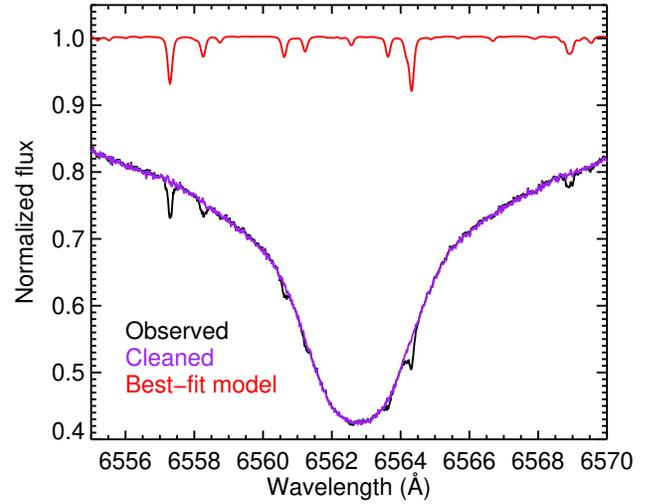}
   \figcaption{Example of the telluric removal procedure near \halpha\ for
   a single WASP-33 exposure. The best-fit telluric model (red spectrum) removes the 
   observed telluric lines down to the noise level.
\label{fig:tellrem}}

\end{figure}

There are numerous telluric H$_2$O lines in the spectrum near \halpha\ which,
if not accounted for, can contribute noise and extraneous features to the
transmission spectrum. We used the telluric modeling procedure \texttt{Molecfit}
\citep{kausch15,smette15} to approximate the telluric spectrum observed for the A0 
spectroscopic standard star HD 89239. We then fit the telluric model to each 
individual WASP-33 spectrum by scaling and shifting the telluric model to match 
the observed telluric line depths. We then divide the best-fit model out of 
the stellar spectrum to produce a cleaned region surrounding \halpha. We show 
a typical telluric model fit and removal in \autoref{fig:tellrem}. The telluric 
lines are removed down to the noise level in an individual spectrum.

\section{Transmission spectrum analysis}
\label{sec:tspecs}

Here we focus on the identification of \halpha\ and H$\beta$ absorption in the atmosphere
of WASP-33 b. The transmission spectrum extraction is complicated by two factors:
1. WASP-33 is a known $\delta$-Scuti star and the spectral lines can be distorted by periodic
pulsations in velocity space; 2. WASP-33 b is nodally precessing and thus the transit
chord is variable in time. We discuss our treatment of both in the following
subsections. 

\subsection{Determining the transit chord}
\label{sec:tchord}

WASP-33 b is nodally precessing at a rate of $d\Omega/dt = 0.4269 \pm
0.0051$ deg yr$^{-1}$ \citep{iorio11,johnson15,watanabe20}
which results in a variable transit chord across the star. Thus in order to
accurately correct the transmission spectrum for the effects of the
occulted stellar surface (i.e., center-to-limb variations and the Rossiter-McLaughlin
effect), up-to-date transit parameters need to be derived from the most
current transit data, ideally the data from which the transmission spectrum
is being extracted. 

We adopt an analytical framework to model the Doppler tomography (DT) signal. In this 
framework, the planet DT signal is a Gaussian perturbation of the stellar line
profile~\citep{Hirano2011}. The line profile extraction and DT modeling are described in detail
in~\citet{Wang2018}. The modeling parameters are the impact parameter, 
projected spin-orbit alignment angle, projected rotational velocity, quadratic limb 
darkening parameters, planet-star radius ratio, systemic velocity, and the mid-transit time.    

\begin{figure*}[htbp]
   \centering
   \includegraphics[scale=.42,clip,trim=0mm 0mm 0mm 0mm,angle=0]{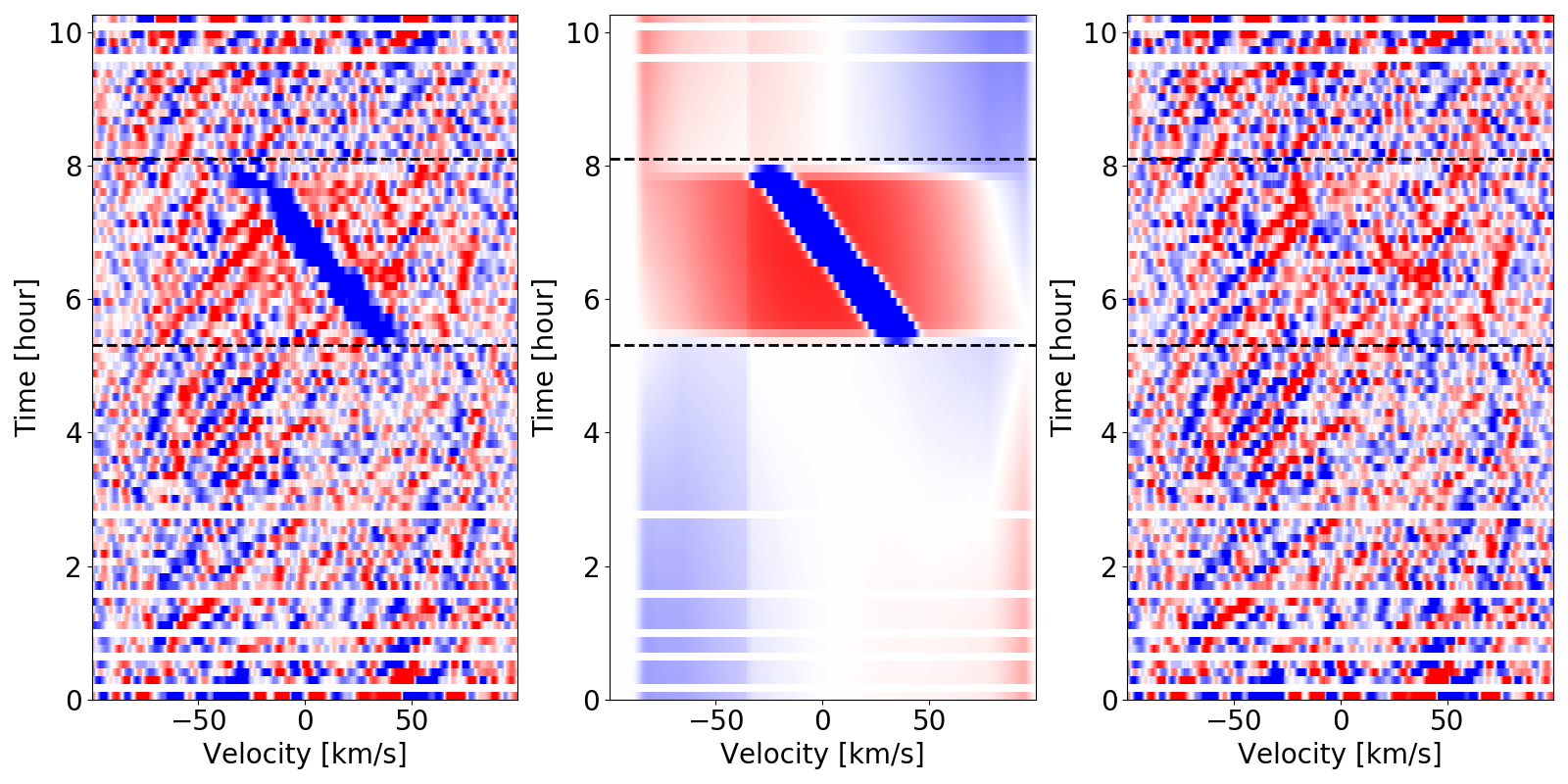}
   \figcaption{{\bf{Left}}: Fourier-filtered residual map after subtracting the median line profile.
   The planet ``Doppler shadow" is the diagonal blue track running from the bottom-right to the
   top-left. Dashed lines mark the first and last transit contact points. Gaps are due to interpolation onto a time array with a fixed interval. {\bf{Middle}}: Modeled DT signal. The overall gradient along the time axis is due to the shift of stellar radial velocity. The enhanced red background during transit is due to the fact that we set the line profile normalization to unity for all line profiles.  {\bf{Right}}: difference of the two maps on 
   the left and middle panel. 
\label{fig:dtmap}}

\end{figure*}

One difficulty in extracting line profiles for WASP-33 is its pulsation as a $\delta$-Scuti 
variable star. We apply a customized Fourier filter to remove the pulsation signal 
owing to the fact that the pulsation signal and the DT signal are almost orthogonal 
to each other. This is evident in the left panel of \autoref{fig:dtmap} where there is still residual 
pulsation signal. Therefore, the two signals can be easily separated in Fourier space. A similar
strategy was adopted in previous works~\citep{johnson15,watanabe20}. The difference is that we 
use a series of inverse normalized Gaussian profiles to mask out the strongest pulsation frequencies in the 
Fourier space until the DT signal stands out clearly. This approach minimizes the ``ringing" 
effect as seen in previous works thanks to the apodizing Gaussian profiles. 

We give the modeled transit parameters from our analysis in \autoref{tab:pars}.
We display maps of the DT signal, the associated model, and the model residuals in
\autoref{fig:dtmap}. We adopt the mid-transit time, spin-orbit
alignment angle, transit duration, and system velocity from our DT analysis given their 
importance in creating the transmission spectrum and the fact that these values
change over time due to the planet's precession. All of the other WASP-33 system 
parameters are taken from the literature. 

\subsection{Constructing the transmission spectra}
\label{sec:construct}

We calculate the individual transmission spectra by dividing each stellar
spectrum by the mean out-of-transit stellar spectrum, which we will refer to
as the ``master-out''. All of the out-of-transit
spectra have the same signal-to-noise due to the use of the photon counter. 
Thus we do not exclude any spectra from the master-out spectrum
based on quality. We do, however, only use spectra with observations
midpoint times of $\lvert T_i \rvert > 2.0$
hours in order to avoid using spectra near the transit as out-of-transit
comparisons. We determine the time from mid-transit for each exposure
using the values of $T_0$ and $T_{14}$ derived in \autoref{sec:tchord}. 
The final master-out is composed of 46 out-of-transit spectra.

Each in-transit transmission spectrum needs to be corrected for the distortions
caused by the occulted portion of the stellar disk during that observation.
The primary effects are center-to-limb variations (CLVs) in the spectral line and
the Rossiter-McLaughlin (RM) distortion caused by occultation of a piece of
the rotating stellar disk \citep[e.g.,][]{czesla15,yan17,cauley19}. To model
these effects we follow the same procedure presented in \citet{yan18} and
\citet{cauley18,cauley19}. For each spectral line of interest, in this
case \halpha\ and \hbeta\, we create synthetic spectra using \texttt{Spectroscopy Made Easy}
\citep{piskunov17} and the stellar parameters in \autoref{tab:pars}. We generate spectra
at 25 different $\mu$-angles to account for CLVs in the spectral lines.

We then generate a grid representing the stellar disk. Each element in the 
grid has dimensions $0.01 R_* \times 0.01 R_*$. We assign a spectrum to each
grid point by shifting that spectrum according to the local rotational velocity
of that grid location and applying the wavelength-dependent limb-darkening
derived from the synthetic spectra calculated at the various $\mu$-angles.
The synthetic out-of-transit spectrum is then the sum of the spectra
for all of the grid points on the stellar disk. We use \texttt{EXOFAST} 
\citep{eastman13} to calculate the planet positions on the stellar disk 
to generate model CLV+RM profiles for the midpoint of each exposure. 
We divide these model profiles out of the transmission spectra
to remove the CLV+RM features. It is worth highlighting that the CLV+RM
profiles only interfere with the planet's absorption signature when 
the absorption profile overlaps the CLV+RM profile in velocity space.
Thus if these signatures are not modeled explicitly then it is acceptable
to simply throw out the in-transit exposures for which the planet's
line-of-sight velocity intersects the local occulted rotational velocity
\citep[e.g.,][]{ehrenreich20,wyttenbach20}.

\begin{figure*}[htbp]
   \centering
   \includegraphics[scale=.68,clip,trim=5mm 5mm 5mm 15mm,angle=0]{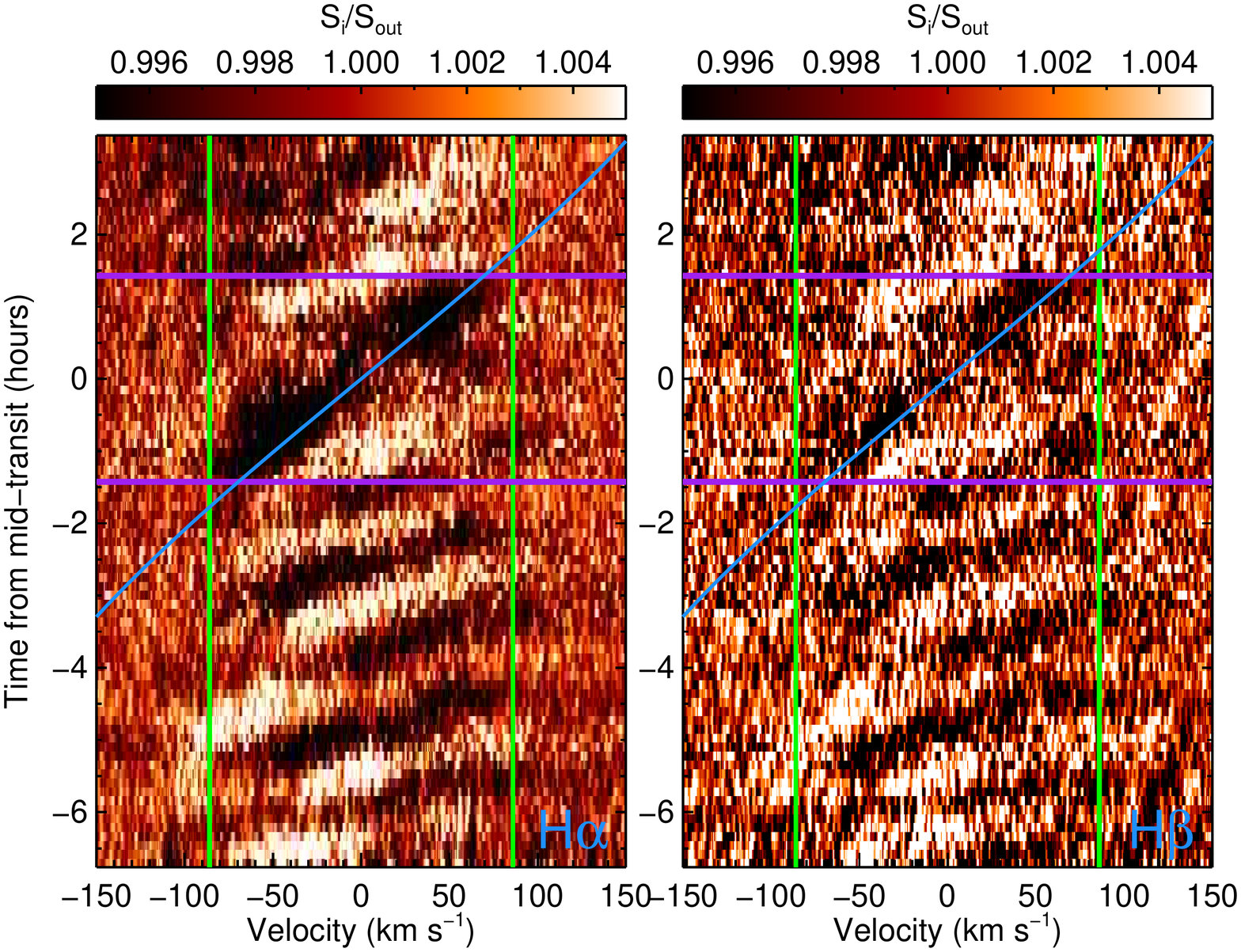}
   \figcaption{Spectral map of the \halpha\ and \hbeta\ transmission spectra in the
   stellar rest frame for the entire night.
   The spectra have been interpolated onto an evenly spaced time vector for display purposes
   which produces some of the smearing near the beginning and end of the night when
   exposures were longer on average.
   The transit contact points T$_1$ and T$_4$ are shown with horizontal purple lines.
   The star's $\pm v$sin$i$ value is marked with the vertical green lines. The planet's
   line-of-sight velocity is shown with the blue line. There is a clear \halpha\ 
   signature which moves along the planet's velocity for the duration of the transit.
   The \hbeta\ absorption is weaker but still present at the expected velocities.
   Note the pulsation stripes visible in the pre-transit data in both lines.
\label{fig:specmap}}

\end{figure*}

We do not consider an increase in magnitude of the CLV+RM profiles
due to an increased effective planet radius in a given spectral line, as 
suggested by \citet{yan18}. There are two reasons for this. The first is
that we see no consistent evidence at \halpha\ or \hbeta\ for an amplified CLV+RM
effect, although a few transmission spectra exhibit larger than expected
signals. The second reason is that absorption by the planet's atmosphere
only increases the magnitude of the CLV+RM profile when the line-of-sight
velocity of the planet overlaps significantly with the local occulted rotation 
velocity. This is because the star effectively ``sees" the larger planetary
radius at the velocity in the spectrum at which the atoms are absorbing.
As we will discuss, the exposures during which the planet's velocity
overlaps the local stellar rotational velocity (middle portion of the
transit in \autoref{fig:specmap}) show anomalously weak
absorption and are not included in much of our analysis. Thus 
we do not consider the enhanced CLV+RM effect since the exposures 
which should most strongly be affected show marginal planetary absorption
and are excluded from the average transmission spectrum
calculation.

We show the spectral time-series maps of the \halpha\ and \hbeta\ transmission
spectrum, S$_i$/S$_\text{out}$, in \autoref{fig:specmap}. The stellar pulsation
signal is visible in the pre-transit data as the dark and light stripes
that extend from approximately $-v$sin$i$ to $+v$sin$i$, which are marked
with the vertical green lines. We show the transit start and end times $T_1$ and 
$T_4$ with the horizontal purple lines. We also show the planet's line-of-sight
velocity, calculated using the value of $K_\text{p}$ from \autoref{tab:pars},
with the blue line. Absorption in both \halpha\ and \hbeta\ is visible during
the transit and closely tracks the planet's velocity. There is a noticeable
lack of absorption during the central portion of the transit. Although the
absorption is expected to be weaker here due to the planet's velocity 
moving across the deepest section of the local stellar \halpha\ and \hbeta\ 
lines, our models, which we discuss in the next section, cannot account
for a complete absence of absorption. We note that a similar effect is
seen by \citet{casasayas19} for the UHJ KELT-20 b, which orbits an
A2 star. 

\begin{figure*}[htbp]
   \centering
   \includegraphics[scale=.68,clip,trim=5mm 25mm 10mm 45mm,angle=0]{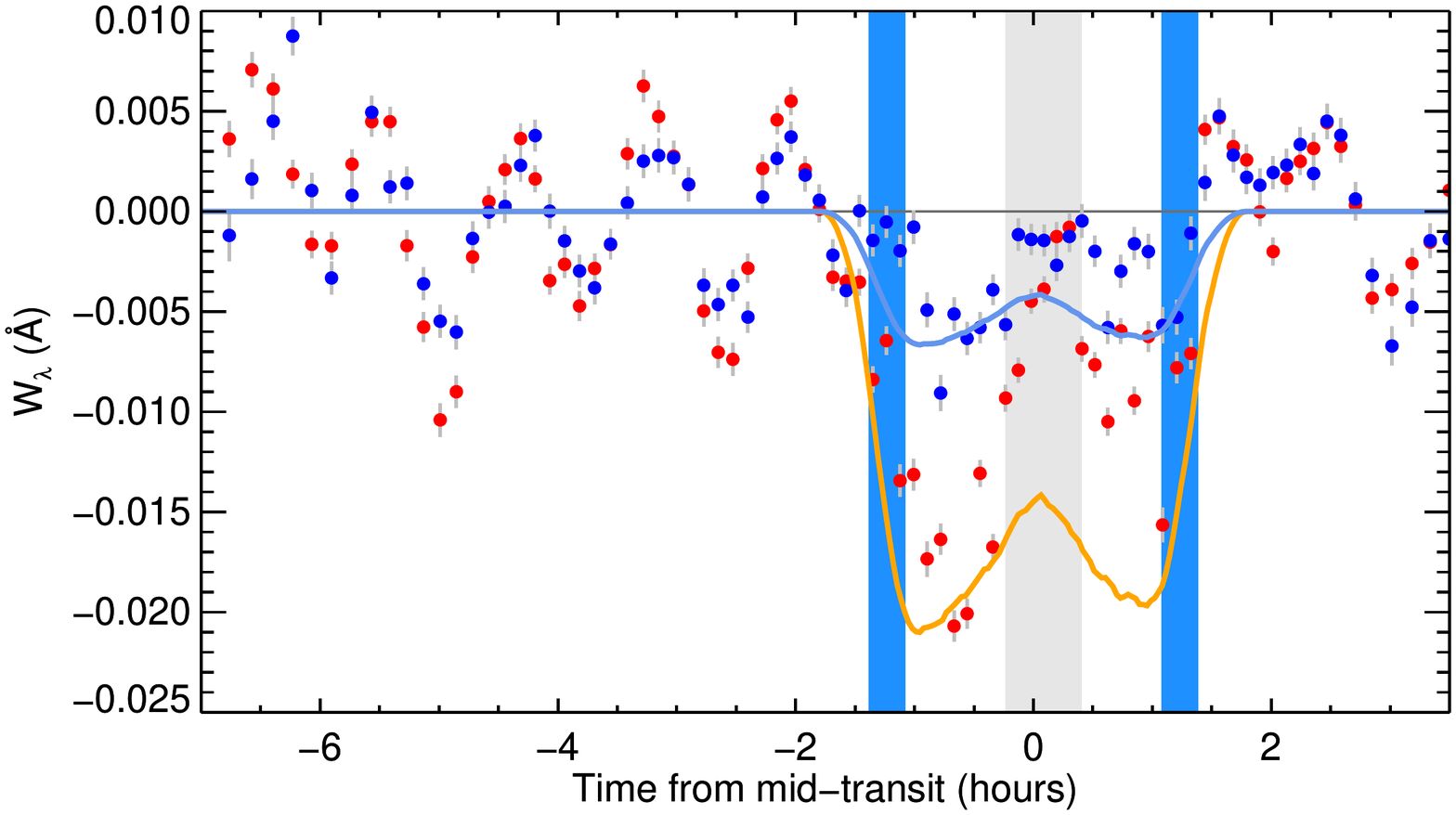}
   \figcaption{Equivalent width time-series of the individual Balmer line transmission 
   spectra. The \halpha\ data is plotted with red circles and the \hbeta\ data is
   shown with blue circles. The ingress and egress portions of the transit are
   the light blue filled regions. The filled gray region represents where the planet's
   absorption signal overlaps with the local RM signal from the stellar surface.
   The absorption models (solid blue and orange lines) are calculated using 
   the best-fit parameters from the transmission spectrum fitting in \autoref{sec:modresults}. 
   There is asymmetry in the transit with the first half absorption being much stronger 
   than the second half. Note the pulsation signal that is visible in the pre-transit data.
\label{fig:weq}}

\end{figure*}

The same anomaly is present in \autoref{fig:weq}, which shows the equivalent width of the
Balmer line transmission spectra as a function of time, where the absorption 
disappears entirely near mid-transit and is weaker during the second half 
of the transit compared with the first half. It's possible that
the stellar photosphere models calculated with \texttt{SME} do not
accurately reproduce the intrinsic Balmer line depths and thus the
modeled CLV+RM signals are underestimated for these portions of
the transit. Another possible explanation for
the weaker than expected absorption is the overlap of the planet's absorption
profile with a pulse profile. Given the pulse model results in the next section,
it seems unlikely that a pulse profile could entirely mask the absorption
during the second half of the transit. We defer a more detailed exploration
of the interplay between the pulse profiles and absorption depths to future
work.

\begin{figure}[htbp]
   \centering
   \includegraphics[scale=.45,clip,trim=45mm 15mm 5mm 20mm,angle=0]{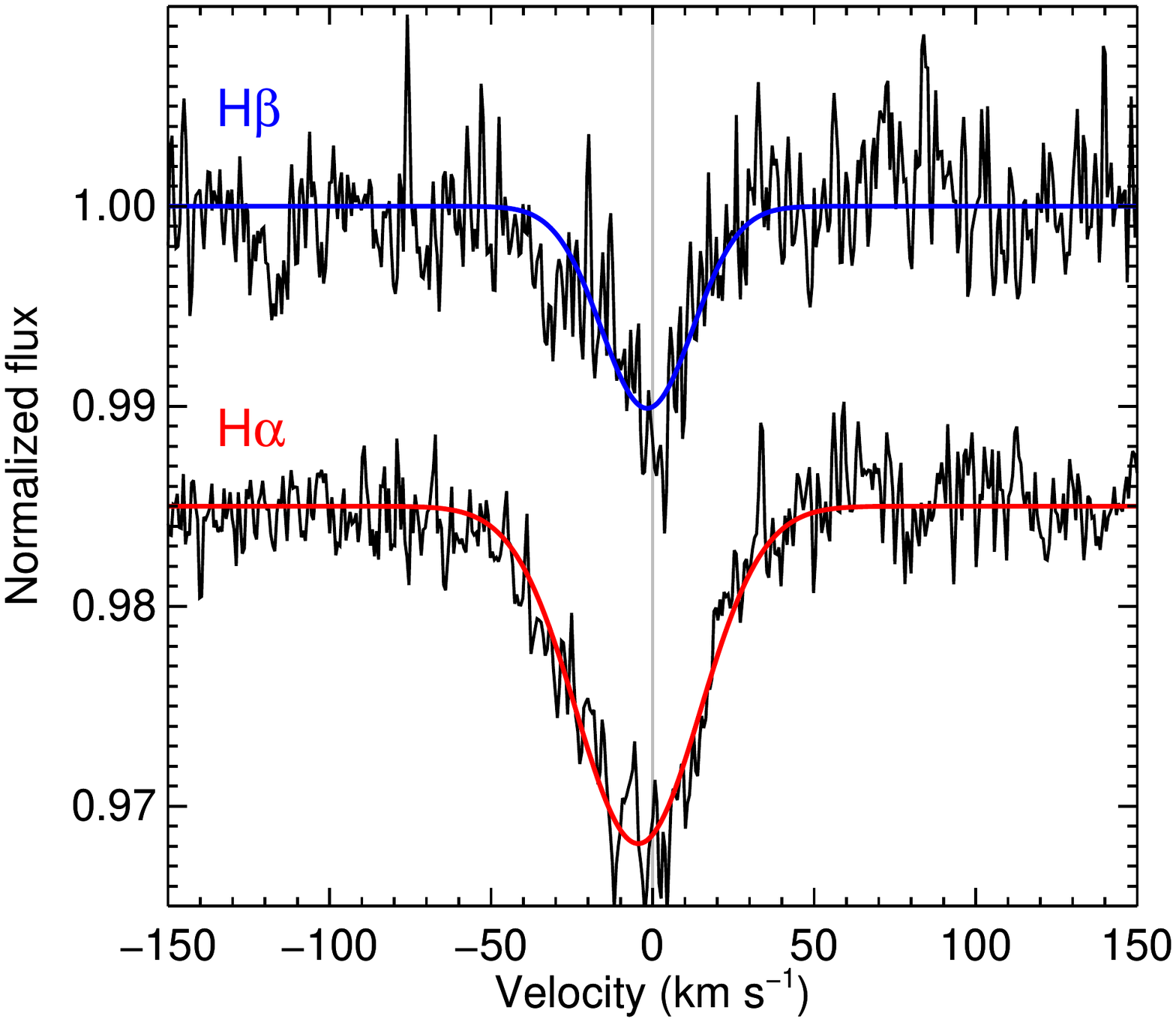}
   \figcaption{Average transmission spectra for H$\beta$ (top) and H$\alpha$ (bottom)
   with best-fit Gaussian profiles over-plotted in blue and red, respectively.
   There is a significant blue-shift in both lines, which we discuss in
   \autoref{sec:modresults}. The profiles have not been corrected for possible
   contamination by the stellar pulsations.
\label{fig:tspecav}}

\end{figure}

We present the average in-transit Balmer line transmission spectra in
\autoref{fig:tspecav}. We only include the spectra taken between 
$-1.0\ \text{hours} \leq t_i \leq -0.56\ \text{hours}$ in the average due to
the lack of absorption in the middle portion of the transit and the
weaker absorption during the second half of the transit. This resulted in
a total of 5 in-transit spectra. Thus it is important to highlight
that our measurements of the average in-transit absorption are biased
towards the stronger absorption during the first half of the transit. 
Before averaging we shift the selected spectra
into the rest-frame of the planet using the orbital and transit parameters
from \autoref{tab:pars}. We include plots of the individual transmission
spectra in the planet's rest frame in \autoref{app:appendix}. We list Gaussian 
fit parameters to the average transmission spectra in \autoref{tab:gfits}. 
Note that the measured blue-shifted velocity offset could be partially due to the 
uncertainty in the transit midpoint timing (see \autoref{sec:midtran}).

We measure the absorption in the average transmission spectra using the
equivalent width integrated from $-100$ km s$^{-1}$ to $+100$ km s$^{-1}$.
These values are listed in column 5 in \autoref{tab:gfits}: the \halpha\ 
absorption is detected at $\approx 44\sigma$ and the \hbeta\ absorption
is detected at $\approx 30\sigma$. The absorption is significant and represents
the first detection of the Balmer lines in WASP-33 b's atmosphere\footnote{See \citet{yan20a}
for a contemporaneous measurement that we became aware of while this manuscript
was under review.}. We note that the maximum peak-to-valley amplitude of
the pre-transit pulsation signal is $0.95\%$ (see \autoref{sec:pulse}) which likely
biases the reported Gaussian parameters. It is interesting to note that \citet{yan20a}
report line depths of $\approx 1.0\%$ and $\approx 0.5\%$ for H$\alpha$ and H$\beta$,
respectively, which is similar to our measured line depths if the maximum 
half-amplitude of the pulse signal is subtracted from the observed line profiles.
We explore the pulse contamination in more detail in \autoref{sec:pulsesim}.

\begin{deluxetable}{lcccc}
\tablecaption{Balmer line Gaussian fit parameters and absorption measurements\label{tab:gfits}}
\tablehead{\colhead{}&\colhead{Contrast}&\colhead{FWHM}&\colhead{$v_0$}&\colhead{W$_\lambda$}\\
\colhead{Line}&\colhead{(\%)}&\colhead{(km s$^{-1}$)}&\colhead{(km s$^{-1}$)}&\colhead{(m\AA)}}
\colnumbers
\tabletypesize{\scriptsize}
\startdata
H$\alpha$ & 1.68$^{+0.02}_{-0.02}$ & 45.4$^{+0.7}_{-0.8}$ & $-4.5^{+0.3}_{-0.3}$ & 17.5$\pm 0.4$\\
H$\beta$  & 1.02$^{+0.05}_{-0.05}$ & 33.0$^{+2.5}_{-2.4}$ & $-1.7^{+0.8}_{-0.8}$ & 12.2$\pm 0.4$\\
\enddata
\end{deluxetable}

\subsection{Stellar pulsations and their effect on the transmission spectrum}
\label{sec:pulse}

WASP-33 is a well known pulsator \citep[e.g.,][]{collier10,herrero11,vonessen14,vonessen20} with
pulsation periods of $\approx 1$ hour. Pulsations in spectroscopic data can be
seen as bumps that, in WASP-33's case, move between $-v$sin$i$
and $+v$sin$i$ in velocity space across the line profile as a function of 
time \citep[e.g.,][]{johnson15}. In our Balmer line time series this is most clearly seen 
in the pre-transit spectra in \autoref{fig:specmap} as the diagonal bright and dark 
bands. Since the pulsation features overlap with the planet's line-of-sight 
velocity and have periods on the order of the transit duration, it is possible 
for the pulsations to distort or contaminate the transmission spectrum.

In order to attempt to predict the in-transit pulse signal and remove it from
the transmission spectra, we modeled the \halpha\ pulses as two periodic pairs of Gaussians 
where each pair has a positive and negative amplitude Gaussian with a fixed
velocity separation. The observed H$\beta$ pulse profiles are, in general, the same as
the \halpha\ pulses to within the flux uncertainties. This also means the H$\beta$ line 
profiles are more strongly affected given their smaller absorption depths. 
The model also includes the FWHM of the pulses, the
minimum and maximum velocities of the pulses, and the rate at which the
pulses traverse the stellar disk. We use a custom MCMC routine based on the
algorithm in \citet{goodman10} \citep[see also ][]{foreman13} to find the
maximum likelihood fit for the pulse model. We assume uniform priors for
all parameters and run the MCMC chains for 10$^5$ steps with 10$^2$ walkers
per chain. We choose the most-likely parameters as the median values of the 
marginalized posterior distributions. 

\begin{figure*}[htbp]
   \centering
   \includegraphics[scale=.65,clip,trim=5mm 5mm 5mm 15mm,angle=0]{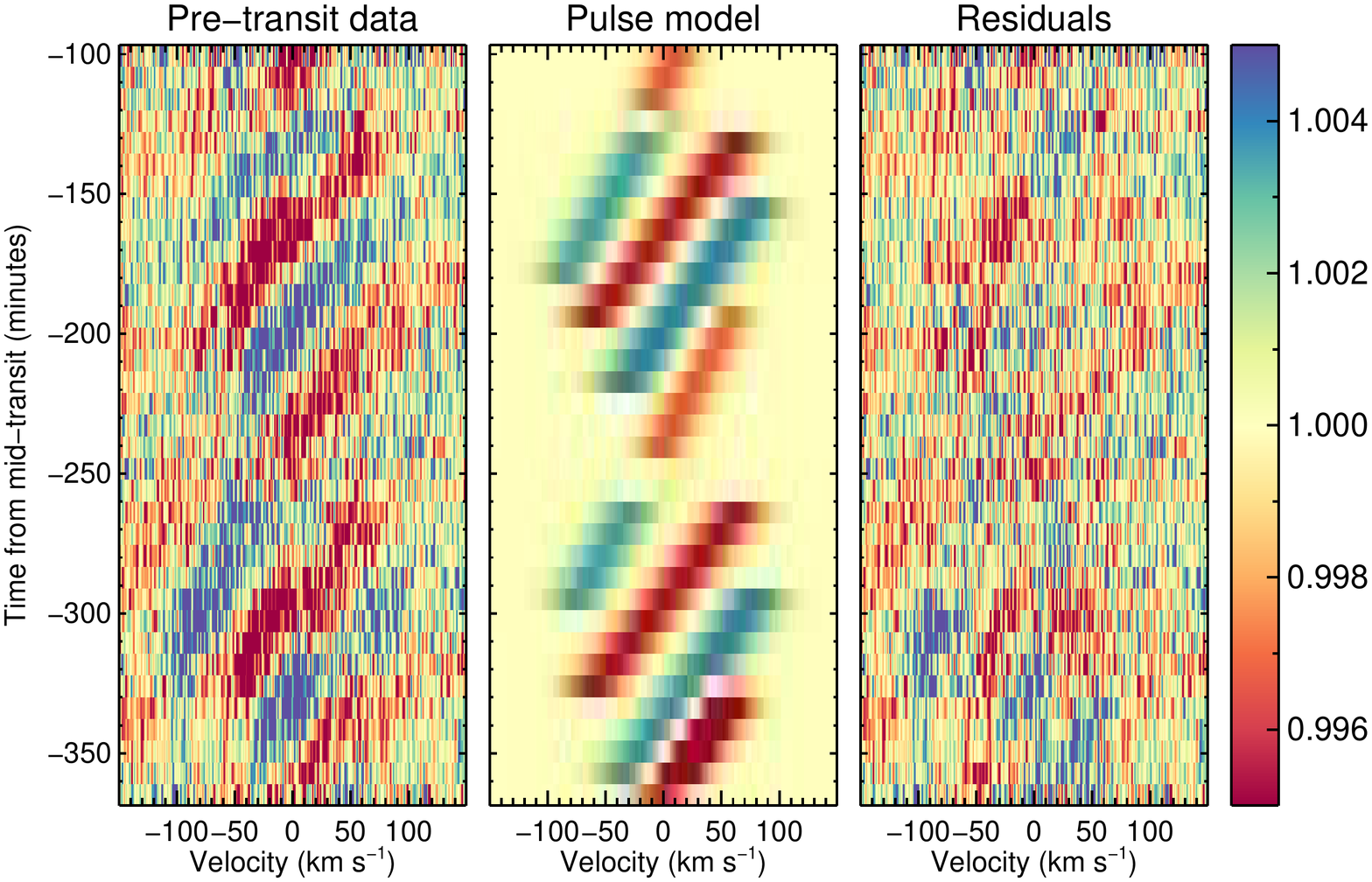}
   \figcaption{Maps of the pre-transit pulse signal. The pre-transit data is in
   the left panel and the best-fit pulse model is shown on the right. The pulse model
   approximately reproduces the structure of the pulses but some features are 
   visible in the residual map.
\label{fig:hapulse}}

\end{figure*}

\begin{figure*}[htbp]
   \centering
   \includegraphics[scale=.7,clip,trim=0mm 5mm 5mm 15mm,angle=0]{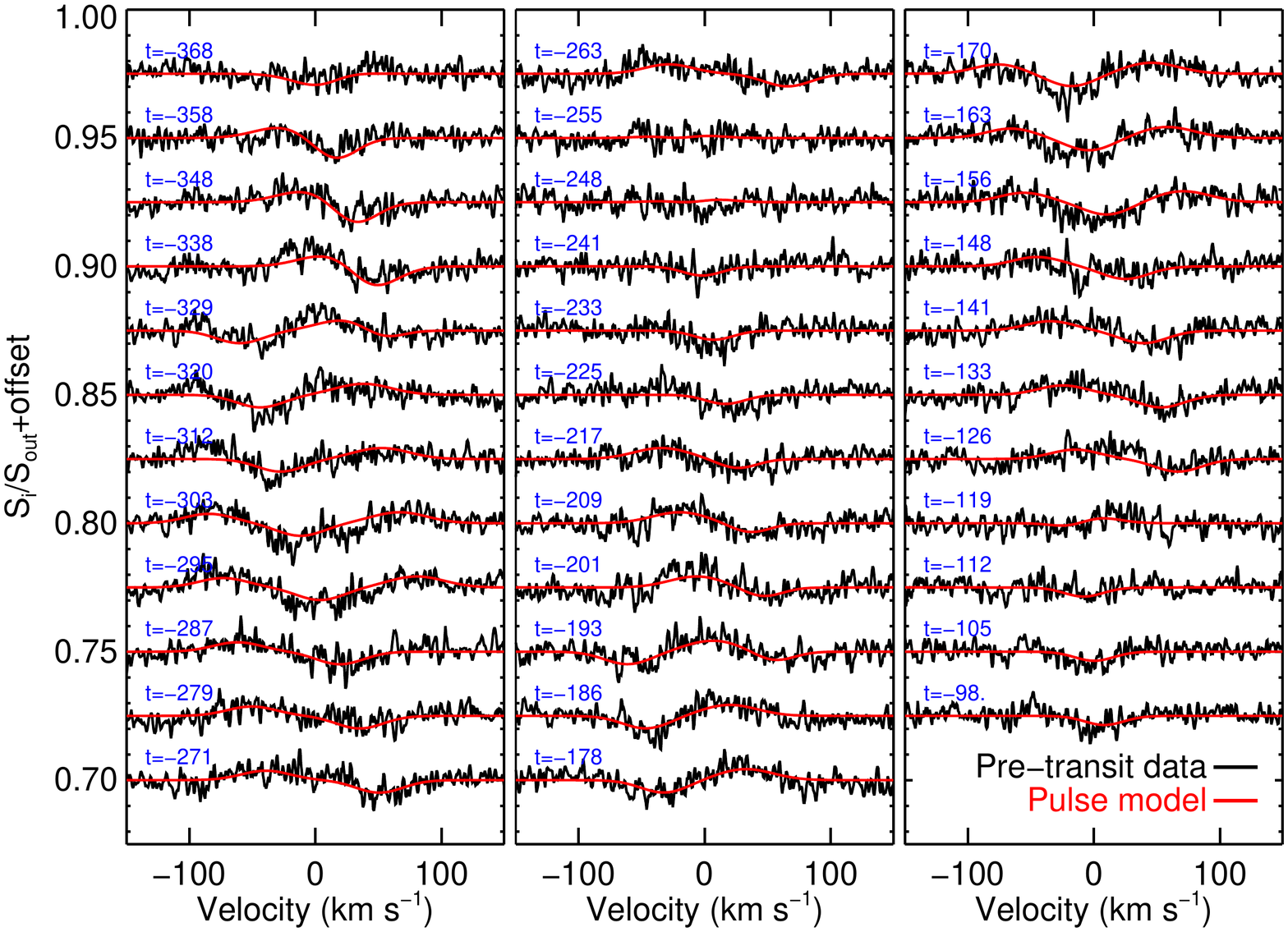}
   \figcaption{Individual pre-transit spectra (black) and the most-likely pulse model
   spectra (red) for \halpha. The time from mid-transit in hours is given in blue.
   Overall, the pulse model provides a good approximation to the individual pulse
   spectra.
\label{fig:pulsefits}}

\end{figure*}

The most-likely pulse model for \halpha\ is shown in \autoref{fig:hapulse} and
\autoref{fig:pulsefits}. In general, the pulse model is a good approximation
to the observed pulse structure in the pre-transit \halpha\ spectra. However,
there are some noticeable discrepancies, for example, at $t=-368$ minutes
and $t=-170$ minutes where the amplitude and phase of the observed pulse
is not predicted by the model. Thus although the model is useful in
understanding the structure of the pulse spectra we do not consider it
accurate enough to predict the in-transit pulse signal, which is more
difficult to quantify due to its overlap with the planetary features. 
Furthermore, it is unclear exactly what effect the transit itself has
on the pulsation signature, which would not be predicted by our simple
model of the out-of-transit pulsations. Thus we choose to forego removal of the 
in-transit pulse signature given the uncertainties involved and the precise 
properties of the Balmer line transmission spectra that we
derive should considered with this in mind. In
order to better understand how the pulse profiles bias the interpretation
of the transmission spectra, we perform an injection/recovery simulation 
in \autoref{sec:pulsesim} and subsequently examine how
the pulses can affect the velocity centroid analysis in subsubsection 5.2.1.

\section{Atmospheric models with velocity dynamics}
\label{sec:mods}

In order to constrain the atmospheric dynamics responsible for the profile
morphologies and measured velocities in WASP-33 b's extended atmosphere, 
we have developed numerical models that take into account the effects of 
rotation, uniform expansion, day-to-night side winds, and equatorial jets 
on the transmission spectrum. The structure of the models is the same as 
those in \citet{cauley19} and we review them here. Our models are three-dimensional
in the sense that we calculate the columns densities and optical depth
vectors through a chord of the atmosphere using a three-dimensional array
but the radiative transfer is performed on the collapsed 2-D atmosphere. 
Thus we do not include asymmetric geometry effects such as a hotter and
more extended westward planetary limb \citep[e.g.,][]{flowers19}. 

We model the planetary atmosphere on the same discrete grid used to simulate
the stellar CLV+RM profiles. The grid points have size $0.01 R_* \times 0.01 R_*$
which for WASP-33 b equates to $0.087 R_\text{p} \times 0.087 R_\text{p}$. 
The atmosphere is spherically symmetric and of uniform density and is 
parameterized by $r$ (in units of$R_p$), the distance above the optical
planetary radius $R_\text{p}$, and the number density
$n$ (in units of cm$^{-3}$). While the assumption of uniform density is not 
strictly correct, sophisticated models of the $n=2$ electronic level number 
density in hot and ultra-hot Jupiter atmospheres suggests that it varies slowly 
with decreasing pressure in the thermosphere where the Balmer lines form 
\citep{huang17,garcia19}. We initially construct the atmosphere in 3D and 
then collapse the grid into the plane perpendicular to the line-of-sight 
between the observer and the star so that the atmosphere grid is now in units 
of column density (cm$^{-2}$).

We compute the transmission spectrum $S_\text{T}(t)$ at time $t$ in the transit
by looping through the atmosphere and extincting the stellar intensity at each 
atmospheric grid point $I^i_*(t)$ by the optical depth $\tau_i$ through the same 
grid point. We do not consider multiple scatterings in the atmosphere. We also 
sum the stellar grid points that are not occulted by the planet nor absorbed by 
the atmosphere and call this spectrum $I_*^\text{in}(t)$. We then add 
$I_*^\text{in}(t)$ to the spectrum absorbed by the planet's atmosphere and 
divide by the out-of-transit spectrum $I_*^\text{out}$, which has been corrected 
for the CLV+RM profile $I_\text{tran}(t)$. Algebraically this can be written as 

\begin{equation}
S_\text{T}(t) = \frac{I_*^\text{in}(t) + \sum_{i} I^i_*(t) e^{-\tau_i}}{I_*^\text{out}-I_\text{tran}(t)}.
\end{equation}

Before the stellar spectrum is extincted we shift the optical depth vector,
which we model as a Voigt profile with Gaussian broadening 
component $v_t$ and Lorentzian component $v_\text{Lor}$,
by the velocity of the local grid point. We assume a uniform day-to-night side
wind speed $v_\text{wind}$ in the atmosphere which requires a single-valued 
velocity shift of the optical depth at each grid point. Note that in reality 
non-uniform wind speeds result in some broadening of the line profile \citep{flowers19} 
since the velocity dispersion is non-zero. However, rotational and thermal broadening 
likely dominate for UHJs like WASP-33 b. 

For the case of rotation we consider the atmosphere to be rigidly rotating so 
that the rotation velocity $v_\text{rot}$ at $R_p$ at the equator scales linearly 
with distance from the rotation axis. For example, for 
$v_\text{rot} = 6.0$ km s$^{-1}$ the velocity at $r = 0.5$ above the equator would 
be 9.0 km s$^{-1}$. As noted by \citet{wyttenbach20},
the line-of-sight velocity along any line perpendicular to the rotation axis
through a 3D rigidly-rotating atmosphere is constant. Thus we can apply the
rotational velocities to the 2D density grid without accounting for the velocities
in the full 3D case. The column density through the atmosphere decreases as a
function of $r$ so the larger velocities contribute least to the rotational
broadening compared with the smaller velocities at lower altitudes. We
assume that the planet's rotation axis is parallel to its orbital angular
momentum vector.

For uniform spherical expansion, the velocity field must be treated in 3D
since each grid point will have a different line-of-sight velocity depending
on the angle between the grid point and the line-of-sight. This results in
a broadening of the optical depth profile along any sight line through the
atmosphere. Because of the need for 3D accounting in this case, we do not
collapse the density array beforehand. Instead, we loop through the atmosphere
and calculate the velocity shifts for each grid point and then sum the
total optical depth along the sight line. Although our models are capable
of modeling expanding atmospheres, the Balmer lines form at pressures
where the upward velocities are expected to be on order of $\approx 1$ km s$^{-1}$
\citep{salz16,wyttenbach20}. Thus uniform expansion cannot account for the velocity
signature seen in \autoref{sec:obsvel} and we do not explore it further.

We also consider the effects of equatorial jets, which contribute to the
broadening of the optical depth profile. In our models we force the 
jets to exist between planetary latitudes of $|\theta|<25^\circ$ and 
we take the jet velocity to be constant as function of latitude and
altitude in the atmosphere. We assume the jet travels in the same direction
as the planet's rotation. The result of including the jet velocity is
increased broadening since the jet speed is only applied to an equatorial 
band in the planet's atmosphere. However, as noted in \autoref{sec:vels}
and demonstrated in \autoref{fig:velexamps}, jets in our model produce
weaker velocity shifts upon ingress or egress when compared with a 
rotating atmosphere of the same velocity. This is due to the jet speed
being constant throughout the atmosphere and the limited latitude contribution
of the jet broadening. 

\subsection{Modeling the Balmer line transmission spectra}
\label{sec:fits}

Before we present the application of the models described in \autoref{sec:mods} 
to the transmission spectra, there are two details which require a more
in-depth discussion given their effects on fitted model parameters.

\subsubsection{Mid-transit time and line-of-sight velocity}
\label{sec:midtran}

Exoplanet transit ephemerides require frequent updates in order to
refine the mid-transit time of future transits. While simultaneous
high-quality photometry is the ideal method for determining transit
parameters for a corresponding spectroscopic transit \citep[e.g.,][]{johnson15},
modeling the spectroscopic transit itself can provide important
constraints on the mid-transit time which are superior to using
out-of-date ephemerides. 

The mid-transit time becomes critically important for measurements
of the velocity offsets in transmission spectra or cross-correlation
measurements: changing the mid-transit time by a few minutes can 
alter the planet's inferred line of sight velocity by $\sim 1-2$ 
km s$^{-1}$ thus shifting the measured velocity of the transmission 
spectrum in the frame of the planet. This can lead to spurious, or
inaccurate, measurements of a day-to-night side wind.

For WASP-33 b we find that the in-transit line-of-sight velocity of the 
planet changes by $\approx 0.8$ km s$^{-1}$ per minute of difference
in the mid-transit time. For example, if the true mid-transit time
differs from our modeled mid-transit time by $\approx 3$ minutes then 
the magnitude of the line-of-sight velocity change will be 
$\approx 2.5$ km s$^{-1}$. 

Although this uncertainty in the planet's line-of-sight velocity
does not affect the transmission spectrum itself, it translates
directly as an uncertainty in the velocity shift of the transmission
spectra into the rest frame of the planet. In turn, this uncertainty
propagates into model parameters which estimate any overall velocity
shifts in the transmission spectrum. Note that the mid-transit
time uncertainty does not affect the structure of the $v_{H\alpha}$
time series since it produces an approximately constant shift for 
all in-transit spectra. For our models the affected parameter
is the wind velocity $v_\text{wind}$. The statistical uncertainty
on our derived mid-transit time (see \autoref{tab:pars}) is
$\approx 1$ minute but this is likely an underestimate; the actual
mid-transit time uncertainty is probably closer to $\approx 3-4$
minutes. Thus in our final determination of $v_\text{wind}$ we 
include an additional uncertainty of 3.0 km s$^{-1}$ to account
for the mid-transit time error. 

\subsubsection{Doppler smearing}
\label{sec:doppsmear}

High-resolution spectroscopic transit observations necessarily have 
exposure times on the order of minutes. The planet, however, is
continuously changing its line-of-sight velocity and position on
the stellar disk. Thus any observation can be approximated by averaging
the instantaneous spectra sampled finely enough from the beginning
to the end of the exposure. The result is an observed spectrum that
has been broadened, or blurred, by the motion of the planet during the exposure,
where longer exposures result in more broadening \citep{ridden16,wyttenbach20}. 
This is critical for transmission spectrum modeling since most models 
tend to simulate the spectrum at a single time, either at mid-transit or at the 
average mid-exposure time for the spectra which have been averaged
to create the transmission spectrum. If the broadening is non-negligible
models must take this effect into account by either including it explicitly, 
which is computationally expensive, or by approximating the effect
with an applied broadening function \citep[e.g.,][]{wyttenbach20}. 

\begin{figure*}[htbp]
   \centering
   \includegraphics[scale=.7,clip,trim=8mm 20mm 8mm 65mm,angle=0]{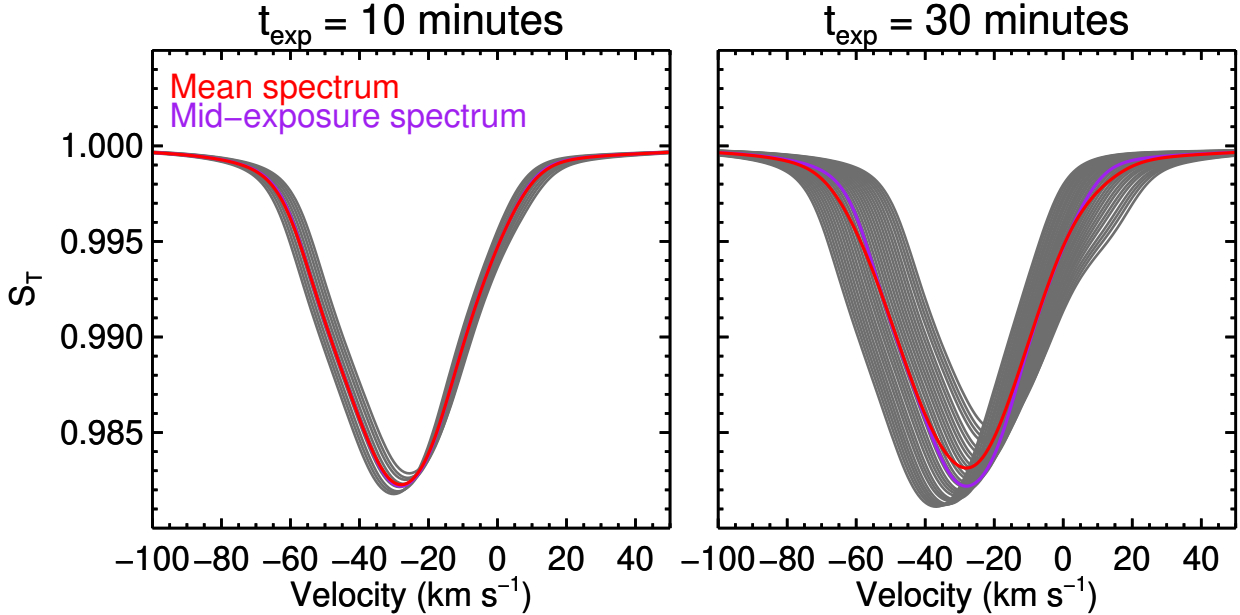}
   \figcaption{Examples of Doppler smearing for WASP-33 b. The left panel
   shows the case of a 10-minute exposure and the right panel shows a
   30-minute exposure. For the 10-minute case there is very little smearing,
   i.e., the mean spectrum of the exposure very closely approximates the
   instantaneous spectrum at the time of mid-exposure. In the 30-minute
   example, however, the smearing effect is noticeable: the mean spectrum
   is broadened relative to the mid-exposure spectrum.
\label{fig:doppsmear}}

\end{figure*}

The question arises: what is the maximum exposure time for a planet
that results in negligible smearing in the observed transmission
spectrum? Using our models we have tested the Doppler smearing effect
for WASP-33 b for an observation with mid-exposure time of $t = -30$ minutes
from mid-transit. We tested multiple exposure lengths from 5 minutes up
to 45 minutes and generated instantaneous spectra at 1-minute intervals
throughout the exposure. Two examples are shown in \autoref{fig:doppsmear} for
exposure duration of $t_\text{exp}=10$ minutes (left panel) and 
$t_\text{exp}=30$ minutes (right panel). The mean spectrum (red line) 
is calculated by averaging all of the instantaneous spectra (gray lines). 
The mid-exposure spectrum is the instantaneous transmission spectrum
at the mid-point of the exposure. Doppler smearing can be ignored when
the differences between the two spectra are negligible. 

While we do not derive any specific relationships for the magnitude of
the effect as a function of exposure duration, it's clear the Doppler
smearing is essentially absent from the $t_\text{exp}=10$ minute spectrum while it
is beginning to manifest in the $t_\text{exp}=30$ minute case. Since
our longest observations are $\approx 10$ minutes we conclude that Doppler
smearing can be safely ignored in our transmission spectrum modeling
described in \autoref{sec:modresults}. This effect should be evaluated
on a case-by-case basis since narrower spectral lines will be more strongly
affected and planets with larger orbital velocities will experience
more dramatic smearing.

\subsubsection{Balmer line model results}
\label{sec:modresults}

We applied the models described in \autoref{sec:mods} to the average \halpha\ and
\hbeta\ transmission spectra shown in \autoref{fig:tspecav}. For each model
iteration the transmission spectra of \hbeta\ and \halpha\ are calculated
simultaneously at the mean $x$ and $y$ locations on the stellar disk of the
exposures included in the transmission spectrum being fit. In other words, we do 
not calculate a spectrum for each exposure and then average them to produce the
model average transmission spectrum but rather calculate a single H$\beta$ and
H$\alpha$ spectrum at the mean transit time of all exposures.

We employ the same maximum-likelihood MCMC routine referenced in \autoref{sec:pulse} 
to find the most-likely model parameters for the Balmer line transmission spectra. We
explore eight scenarios, each a subset of the simplest case which only includes
thermal broadening. The model abbreviations are letters which correspond to
the descriptors Thermal (T), Rotation (R), Wind (W), and Jet (J). We also
test a special case, Model TWJ+R$_\text{tl}$ which fixes the rotational 
velocity at the tidally locked value of $7.1$ km s$^{-1}$ but allows the jet velocity to vary. 
We then compare the fit results using the Bayesian information
criterion (BIC). The potential free parameters in the model are number density
$n$, radial extent $r$ above $R_p = 1.0$, thermal broadening $v_t$, Lorentzian 
broadening $v_\text{Lor}$, rotational broadening $v_\text{rot}$, the day-to-night
side wind velocity $v_\text{wind}$, and the equatorial jet velocity $v_\text{jet}$. 
In the models which do not include rotation, a day-to-night side wind,
or a jet the corresponding parameter values are set equal to zero. We assume broad uniform 
priors for all parameter values.

Each MCMC chain is initiated with 100 independent walkers which are each run for 1500 
steps. We eliminate the first 500 steps as burn-in and use the remaining
samples, which we test for convergence using the Gelman-Rubin statistic, to generate 
1D and 2D posteriors for the parameters. We choose the best-fit parameters by
taking the median value of the marginalized 1D posteriors; we calculate 
$1\sigma$ confidence intervals as the 68\% regions around the median values. 

It is important to note that due to the finite size of the pixels in our
model grid, the variable $r$ is discrete and thus the confidence intervals
determined from the marginalized posterior are not strictly accurate. They can be
thought of as lower bounds on the actual confidence interval since MCMC steps
with $r$ values between pixels with values $r_1$ and $r_2$ were rejected so
only values of $r$ up to and including $r_1$ are realized in the posterior.
We adopt a conservative confidence interval of $0.5$ pixels, or $0.045 R_p$,
for the confidence intervals on $r$ which are less than this.

\begin{figure*}[htbp]
   \centering
   \includegraphics[scale=.75,clip,trim=5mm 10mm 35mm 15mm,angle=0]{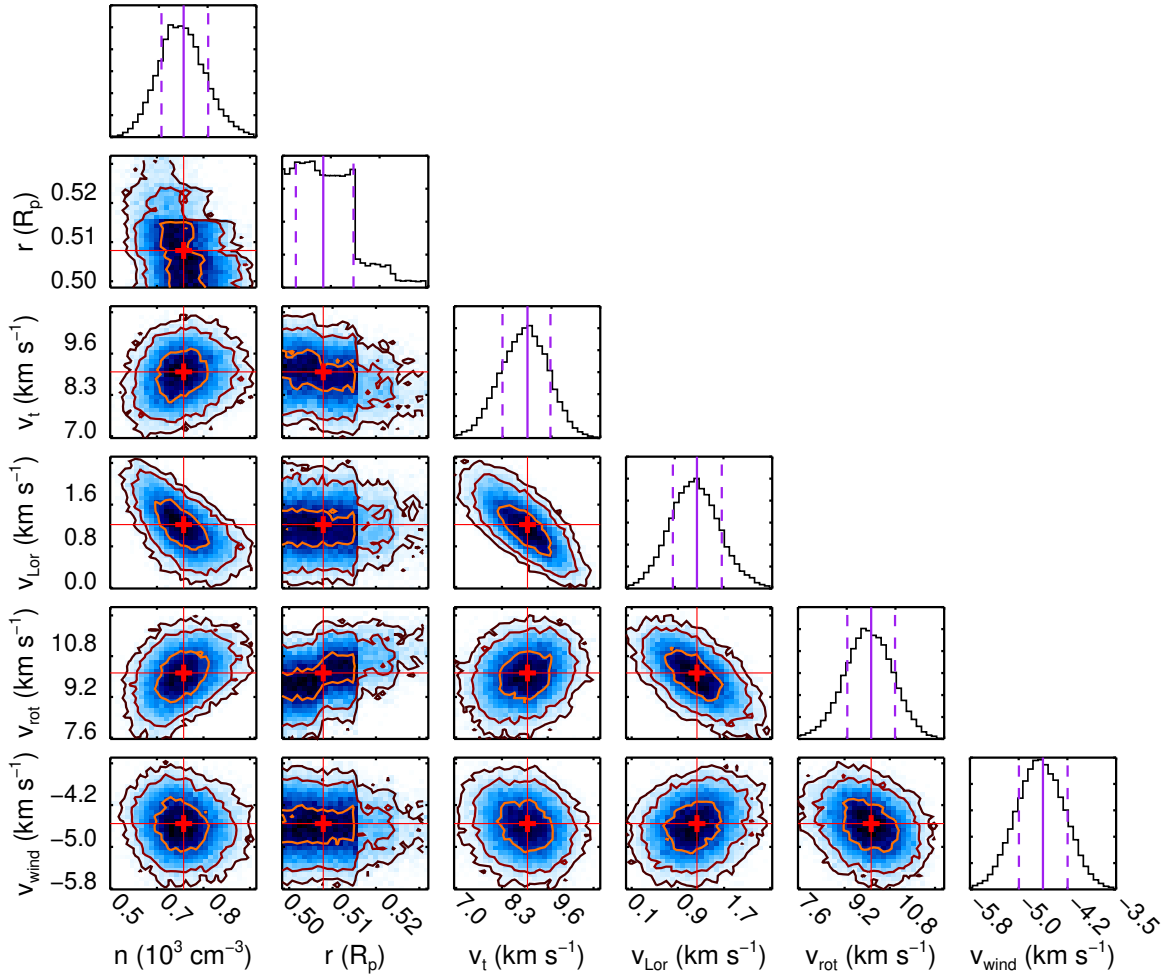}
   \figcaption{Corner plot of the posteriors for the TRW
   model. The marginalized parameter histograms are shown at the top of each
   column and the most-likely parameter value is marked with a vertical purple
   line; the $1\sigma$ confidence intervals are marked with dashed lines. 
   The contours represent the 68\%, 95\%, and 99\% regions of the posteriors.
\label{fig:corner}}

\end{figure*}

\begin{figure*}[!hbt]
   \centering
   \includegraphics[scale=.6,clip,trim=10mm 15mm 5mm 15mm,angle=0]{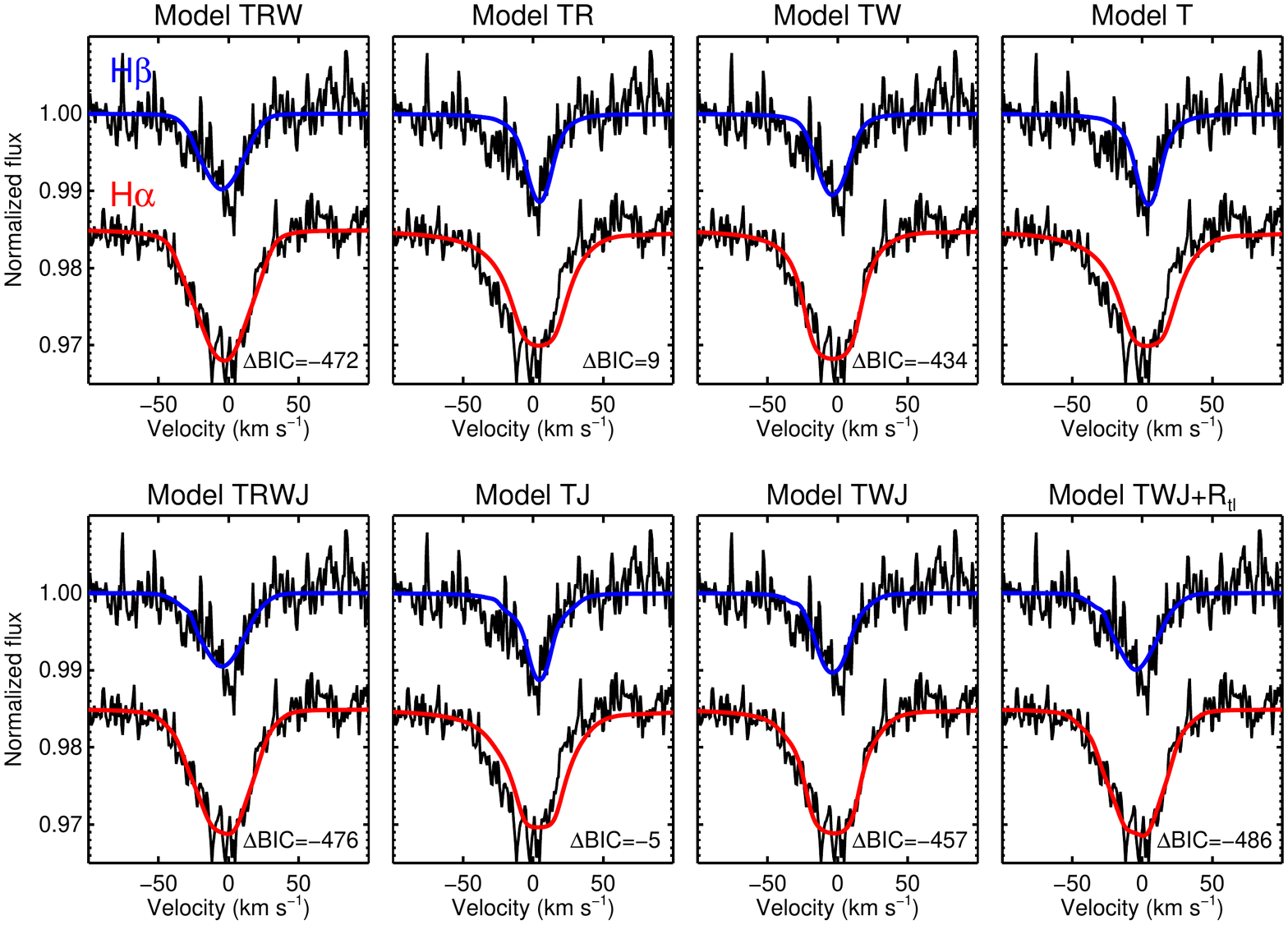}
   \figcaption{Atmospheric model fits (solid colored lines) to the Balmer line 
   transmission spectra (black). The difference between the BIC of each model and
   that of the Thermal model (top right panel, Model T) is given in the lower-right of each plot
   window. The model which includes a day-to-night side wind, rotation, and thermal
   broadening (Model TRW) is strongly preferred over the other models. Although including
   an equatorial jet results in an equally good or better description of the data (Models TRWJ
   and TWJ+R$_\text{tl}$) the jet velocities are non-physical. 
\label{fig:modfits}}

\end{figure*}

We show the corner plot of the 1D and 2D histograms for Model TRW
in \autoref{fig:corner} and the results of the eight model fits in 
\autoref{fig:modfits}. We list the most-likely parameters and their uncertainties in
\autoref{tab:mfits}. The $\Delta$BIC values are relative to the Thermal (T) model. It's 
clear upon visual inspection of the line profile fits that the models which include
the day-to-night side wind are strongly favored over the models with no wind. This
was foreshadowed by the Gaussian fits to the average transmission spectra which
also found significant blue-shifts in the spectra. For the no-wind models the
natural broadening $v_\text{Lor}$ tends to be large since the MCMC is attempting
to find solutions that can match the blue side of the profile. Thus the resulting
$v_\text{Lor}$ values are unrealistic since there is no pressure broadening at the 
densities and pressures where \halpha\ forms in hot planet atmospheres
\citep{huang17,wyttenbach20,turner20}. 

Although the models which include an equatorial jet and a wind (Models TRWJ,
TWJ, and TWJ+R$_\text{tl}$) show similar or lower BIC values, the most-likely
jet velocities are likely too large to be physical (\autoref{tab:mfits}):
all three jet models find a jet velocity of $\approx 19-20$ km s$^{-1}$.
GCMs currently do not explore micro- and nano-bar pressures but there is little
evidence that such large jet velocities are possible even in the extreme
atmosphere of UHJs \citep{flowers19,carone20}. Given the unrealistic jet
values found by our models, we choose to proceed by only considering the
rotation models and their explanatory power for the measured in-transit
transmission spectrum velocities.

Excluding the jet models based on the non-physical jet velocities we obtain, 
the TRW and TW models both provide better descriptions of the Balmer line 
transmission spectra than the purely thermal
model. This is verified by their significantly lower BIC values. Including
rotation gives the best description of the data: the BIC value for the TRW 
model is 30 points lower than the TW model, strong evidence that 
the TRW model is a better approximation to the atmosphere
than the TW model. We conclude that incorporating rotation into
the model line profiles provides a more accurate description of WASP-33 b's
atmospheric physics. Our model suggests that the rotational
velocity probed by the Balmer lines is $v_\text{rot} = 10.1^{+0.8}_{-1.0}$ km s$^{-1}$, 
$\approx 3$ km s$^{-1}$ greater than expected for the case of a tidally locked 
WASP-33 b. We also derive a day-to-night side wind speed of 
$v_\text{wind} = -4.6^{+3.4}_{-3.4}$ km s$^{-1}$ where we have included an 
estimate of the mid-transit time uncertainty in the confidence intervals 
for $v_\text{wind}$. 

\begin{deluxetable*}{lcccccccc}
\tablecaption{Balmer line model fit parameters\label{tab:mfits}}
\tablehead{\colhead{}&\colhead{$n$}&\colhead{$r$}&\colhead{$v_t$}&\colhead{$v_\text{Lor}$}&
\colhead{$v_\text{rot}$}&\colhead{$v_\text{wind}^{\dagger}$}&\colhead{$v_\text{jet}$}&\colhead{}\\
\colhead{Model$^\star$}&\colhead{(cm$^{-3}$)}&\colhead{($R_p$)}&\colhead{(km s$^{-1}$)}&\colhead{(km s$^{-1}$)}&
\colhead{(km s$^{-1}$)}&\colhead{(km s$^{-1}$)}&\colhead{(km s$^{-1}$)}&\colhead{$\Delta$BIC}}
\colnumbers
\startdata
T & 433$^{+139}_{-96}$ & 0.48$^{+0.04}_{-0.04}$ & 3.8$^{+1.1}_{-1.6}$ & 5.2$^{+0.5}_{-0.7}$ & \nodata & \nodata &\nodata & \nodata\\
TW & 541$^{+168}_{-87}$ & 0.50$^{+0.04}_{-0.04}$ & 7.8$^{+0.9}_{-1.1}$ & 3.6$^{+0.7}_{-0.6}$ & \nodata & -3.5$^{+0.5}_{-0.5}$ &\nodata &$-440$\\
TR & 463$^{+117}_{-95}$ & 0.49$^{+0.04}_{-0.04}$ & 4.3$^{+1.1}_{-1.1}$ & 5.0$^{+0.6}_{-0.7}$ & 1.8$^{+3.1}_{-1.5}$ & \nodata &\nodata &$-9$\\
TRW & 808$^{+94}_{-93}$ & 0.51$^{+0.04}_{-0.04}$ & 8.9$^{+0.7}_{-0.7}$ & 1.3$^{+0.5}_{-0.4}$ & 10.1$^{+0.8}_{-1.0}$ & -4.6$^{+0.4}_{-0.4}$ &\nodata & $-477$\\
TJ & 554$^{+150}_{-123}$ & 0.50$^{+0.04}_{-0.04}$ & 4.3$^{+1.1}_{-1.3}$ & 3.7$^{+0.6}_{-0.6}$ & \nodata & \nodata & 20.4$^{+8.0}_{-7.6}$ & -5\\
TWJ & 666$^{+85}_{-77}$ & 0.51$^{+0.04}_{-0.04}$ & 7.3$^{+0.9}_{-0.8}$ & 2.4$^{+0.4}_{-0.4}$ & \nodata & -3.6$^{+0.4}_{-0.4}$ & 19.2$^{+2.3}_{-2.6}$ & $-458$\\
TRWJ & 826$^{+100}_{-99}$ & 0.52$^{+0.04}_{-0.04}$ & 8.7$^{+0.8}_{-0.9}$ & 0.9$^{+0.5}_{-0.5}$ & 7.2$^{+1.2}_{-0.9}$ & -4.7$^{+0.5}_{-0.4}$ & 19.1$^{+3.1}_{-9.3}$ & $-477$\\
TWJ+R$_\text{tl}$ & 777$^{+100}_{-92}$ & 0.55$^{+0.04}_{-0.04}$ & 8.0$^{+0.8}_{-0.8}$ & 0.9$^{+0.4}_{-0.3}$ & 7.1 & -4.7$^{+0.4}_{-0.4}$ & 20.1$^{+2.8}_{-2.4}$ & $-486$\\
\\
Inj/Rec TRW & 1197$^{+108}_{-95}$ & 0.50$^{+0.04}_{-0.04}$ & 9.1$^{+0.4}_{-0.5}$ & 0.4$^{+0.3}_{-0.2}$ & 8.5$^{+0.7}_{-0.7}$ & -3.4$^{+0.3}_{-0.3}$ &\nodata & \nodata\\
\enddata
\tablenotetext{\dagger}{Table values only include uncertainties from the model fitting
and do not take into account the uncertainty in $T_0$ (see \autoref{sec:midtran})}
\tablenotetext{\star}{Abbreviations: T = Thermal, R = Rotation, W = Wind, J = Jet}
\end{deluxetable*}

\subsubsection{Testing atmospheric parameter bias caused by the pulsation signal}
\label{sec:pulsesim}

For certain in-transit exposures the extrema of the pulse profiles overlap
with the planetary absorption profile. In these cases, the pulse profile likely
modifies the shape and depth of the absorption line. Although our pulse model from
\autoref{sec:pulse} is not accurate enough to remove the pulses from the transmission
spectra, we can use the model to estimate the effect that the pulses have on
the retrieved model parameters from \autoref{sec:modresults}.

To this end we perform a simple injection/recovery simulation using the
most-likely parameter values for the TRW model in \autoref{sec:modresults}. We
first calculate the noise in the continuum for each individual transmission 
spectrum used in the average transmission spectrum. We then use this flux uncertainty 
value to generate noise for the model transmission spectra. We compute model
transmission spectra for each of the five in-transit observation times and
add the simulated noise vectors to the spectra. We then compute the predicted
pulse profile for the same observation times and add it to the simulated
transmission spectra. We shift the five spectra into the planetary rest frame,
calculate the average transmission spectrum, and normalized the transmission spectrum
using the continuum outside of the line absorption. Finally, we run the same 
MCMC routine on the simulated data, which include the predicted pulse contributions,
for the TRW scenario.

\begin{figure}[t!]
   \centering
   \includegraphics[scale=.55,clip,trim=55mm 25mm 15mm 45mm,angle=0]{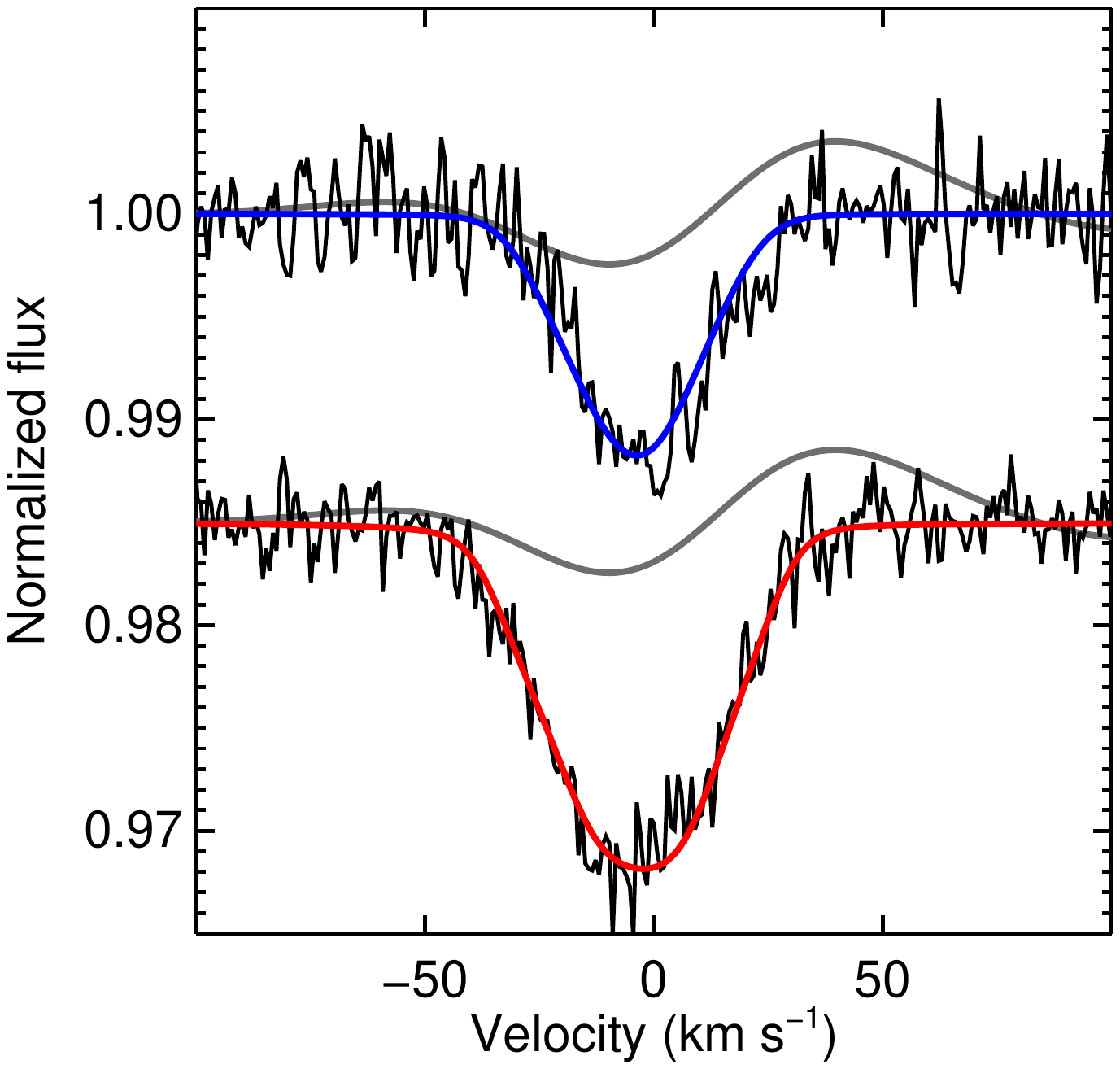}
   \figcaption{Simulated average Balmer line transmission spectra using the
   most-likely TRW model parameters from \autoref{tab:mfits} and the predicted
   pulse contamination from \autoref{sec:pulse}. The best-fit
   model profiles are the red and blue lines. The dark gray lines show the average
   pulse profile from the observations time associated with the spectra used
   to create the transmission spectrum. The most likely injection/recovery
   parameters are similar to the most likely parameters from the observed
   transmission spectrum, suggesting that the pulses do not strongly bias
   the model fits.
\label{fig:injrec}}

\end{figure}

We give the most-likely parameter values for the injection/recovery test
in the last row of \autoref{tab:mfits}. In \autoref{fig:injrec} we show the 
simulated average transmission spectrum, which includes the simulated
pulse contamination, and the best-fit injection/recovery models. We overplot 
the average pulse profile in \autoref{fig:injrec} as a dark gray line to show 
the relative magnitude of the pulse signal and the planetary absorption signal. 

Overall, there are few significant differences
between the true model parameters and the most-likely values from the
injection/recovery test. The exceptions are the density $n$ and the Lorentzian
broadening component $v_\text{Lor}$. The density is sensitive to the ratios
of the line depths: as the H$\alpha$ to H$\beta$ ratio becomes smaller,
the density must increase. Thus increasing the line strengths by the same 
absolute amount, which is what the pulse profile can be seen to do in
\autoref{fig:injrec}, will cause the density to increase. The Lorentzian 
velocity mainly affects the shape of the line wings. The pulse profile in
\autoref{fig:injrec} causes the line depth in the wings to decrease, 
making the absorption profile more Gaussian-like. Thus $v_\text{Lor}$
decreases to account for the shallower line wings. The pulses also appear
to alter the value of $v_\text{wind}$ by $\approx 1$ km s$^{-1}$ but
this is smaller than the uncertainty in the velocity offset caused
by the mid-transit time uncertainty. Thus we don't consider this to
be a problematic difference.

We conclude that although the pulse profiles can alter the shape 
of the transmission spectra, they do not strongly affect the recovered
fit parameters within our modeling framework. We note that we also tested
parameter recovery without the pulse profiles and the retrieved 
values were within $1\sigma$ of the the true values. Thus the differences
between the most-likely parameters found for the observed spectra and
those from the injection/recovery pulse test are due to the pulse alterations
and are not statistical fluctuations. Even if the pulses do not, in general,
strongly affect the parameters recovered from model fitting, our reported
final model parameters should still be viewed with appropriate caution
given the mild contamination that is likely present from the pulses.

\subsection{Velocity measurements of individual \halpha\ transmission spectra}
\label{sec:obsvel}

Due to the low signal-to-noise of the individual \hbeta\ transmission spectra,
we focus our analysis of the velocity centroids on the \halpha\ profiles.
Before calculating $v_{H\alpha}$ we shift the spectra in \autoref{fig:specmap}
into the rest-frame of the planet. Thus any residual velocity signature is due to 
mass motion in the planet's atmosphere or a systematic offset in the calculated
in-transit line-of-sight velocities. 

\begin{figure*}[htbp]
   \centering
   \includegraphics[scale=.60,clip,trim=0mm 5mm 5mm 5mm,angle=0]{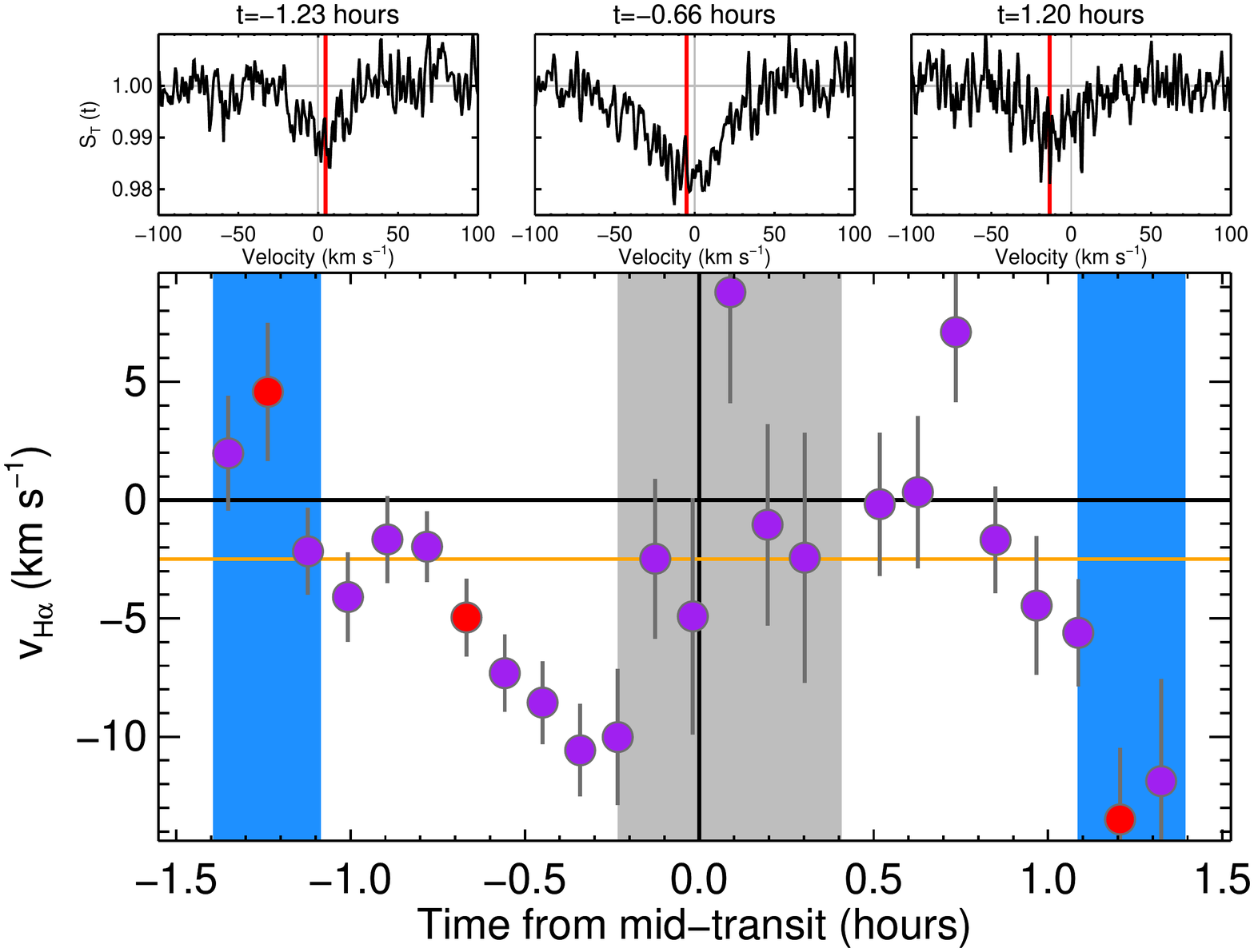}
   \figcaption{Measured $v_{H\alpha}$ values for the in-transit
   \halpha\ transmission spectra. The blue shaded regions show ingress and egress
   times and the gray shaded mark the portion of the transit for which there
   is little or no absorption due to the planet's velocity overlapping with the
   local stellar RM signal. Example transmission spectra are shown above the velocity
   plot where the red plot symbols correspond to the $v_{H\alpha}$ values of
   the example spectra. The orange line marks the mean of the $v_{H\alpha}$ values.
   The best-fit wind velocity from \autoref{sec:fits} is similar to the
   mean $v_{H\alpha}$ value and the structure of the velocities is comparable to
   the rotational velocity examples in \autoref{fig:velexamps}.
\label{fig:havels}}

\end{figure*}

\autoref{fig:havels} shows the in-transit $v_{H\alpha}$ values calculated using 
\autoref{eq:vha} and examples of the \halpha\ transmission spectra during ingress, in-transit,
and egress. The gray shaded region shows where the absorption weakens due to
overlap with the local RM profile; we exclude points in this region since
the absorption is too weak for a reliable measurement. We note that
$v_{H\alpha} = 18.8 \pm. 3.9$ km s$^{-1}$ near $t = 0.4$ hours and is excluded from the 
plot for clarity. This high value is mainly a result of the planet's
absorption profile still overlapping the local stellar rotational velocity and
not being fully corrected by our model CLV+RM profiles. Note that we do not include the
mid-transit time uncertainty in the $v_{H\alpha}$ uncertainties since
this source of error is an offset and would result in the $v_\text{wind}$
taking on a different value but would not increase the uncertainties on
the measurement of $v_{H\alpha}$. 

There is striking similarity in the shape of the $v_{H\alpha}$ 
values when compared with the example rotational models in \autoref{fig:velexamps},
although there is a noticeable offset towards blue-shifted velocities. The
mean value of $v_{H\alpha}$, shown with the horizontal orange line, is
similar to the best-fit wind velocity from \autoref{sec:fits} ($\overline{v_{H\alpha}} = 
-2.4$ km s$^{-1}$ versus $v_\text{wind} = -4.6$ km s$^{-1}$) where
$\overline{v_{H\alpha}}$ is likely skewed towards a smaller blue-shift due to
the large positive values of $v_{H\alpha}$ near $t=0.0$ and $t=0.4$ hours. We note that the
shape of the velocity time series cannot be described by spherically
expanding atmosphere since the ingress and egress velocities in that
case are $\approx 0.0$ km s$^{-1}$. For that reason we do not consider
expansion as an explanatory mechanism for the observations.

\subsubsection{Effect of the pulsation profiles on individual velocity centroids}

\begin{figure}[htbp]
   \centering
   \includegraphics[scale=.35,clip,trim=5mm 5mm 15mm 40mm,angle=0]{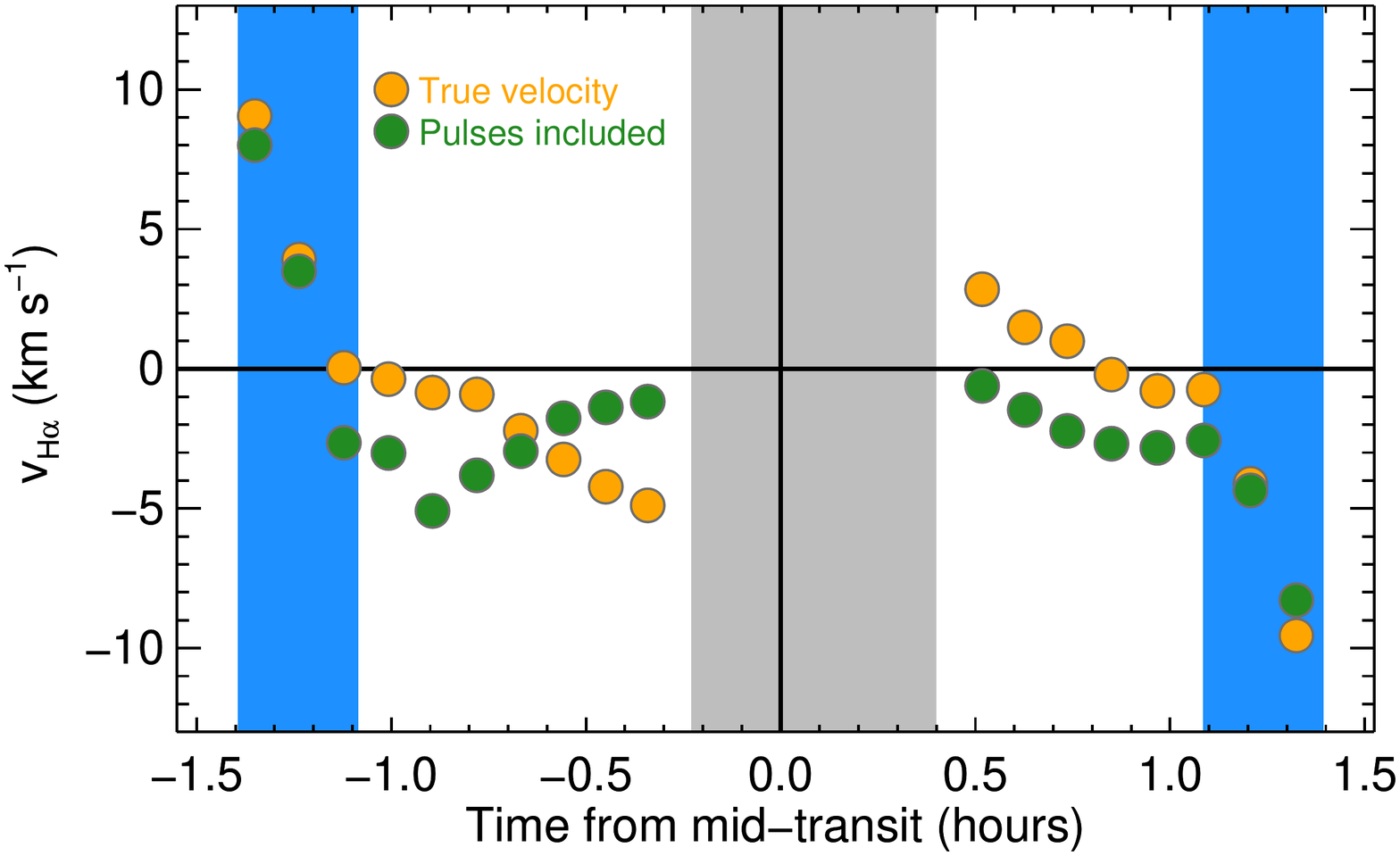}
   \figcaption{Comparison of the true velocity centroids (orange circles) from the 
   simulated data generated using the TRW model most-likely parameters with the velocity
   centroids from the same model profiles which have been modified by the predicted
   pulse profiles (green circles). The pulse-contaminated velocity centroids differ
   from the true velocities by $\approx 2.2$ km s$^{-1}$ on average but the ingress
   and egress velocities are largely unaffected.
\label{fig:velsims}}

\end{figure}

We tested how the predicted pulse profiles affect the $v_{H\alpha}$ 
measurements. We computed the model absorption profiles for each observation
time in \autoref{fig:havels}, except for the CLV+RM contaminated exposures, 
again using the most-likely parameter values from
the TRW model fit. We first calculate the true value of $v_{H\alpha}$ and then
add the predicted pulse profile to the absorption profile. We then recalculate 
$v_{H\alpha}$ for the pulse-modified profile. We show the results in \autoref{fig:velsims}
where the orange circles show the true $v_{H\alpha}$ values and the green circles
represent the pulse-modified $v_{H\alpha}$ values.

On average the pulses shift the true velocities by $\approx 2.2$ km s$^{-1}$ 
but the overall shape of the velocity time series is very similar. It's interesting 
to note that the ingress and egress velocities are less affected than the in-transit 
velocities. This is because the pulses more frequently occur at velocities less than 
the planet's line-of-sight velocity during ingress and egress, which is $\approx \pm 70$ 
km s$^{-1}$. This is clear from the magnitude of the pulse signal seen in \autoref{fig:specmap},
where the largest pulse peaks are generally seen between $\pm 50$ km s$^{-1}$.

\subsection{Comparing the velocity centroids to rotational models}
\label{sec:rotmods}

In \autoref{sec:mods} we derived a rotational velocity for the atmosphere by applying
transmission spectrum models to an average in-transit Balmer line spectrum. As we
noted previously, however, there is some degeneracy between broadening mechanisms
which may not be fully accounted for with our parameterized models. Another test of
the rotational broadening hypothesis is to compare models of varying $v_\text{rot}$
to the in-transit $v_{H\alpha}$ values with the best-fit value of $v_\text{wind}$
subtracted off. If rotation is responsible for the shape of the transmission spectra 
then rotation models should also be able to account for the $v_{H\alpha}$ time series 
in \autoref{fig:havels}. 

\begin{figure*}[htbp]
   \centering
   \includegraphics[scale=.65,clip,trim=0mm 25mm 15mm 30mm,angle=0]{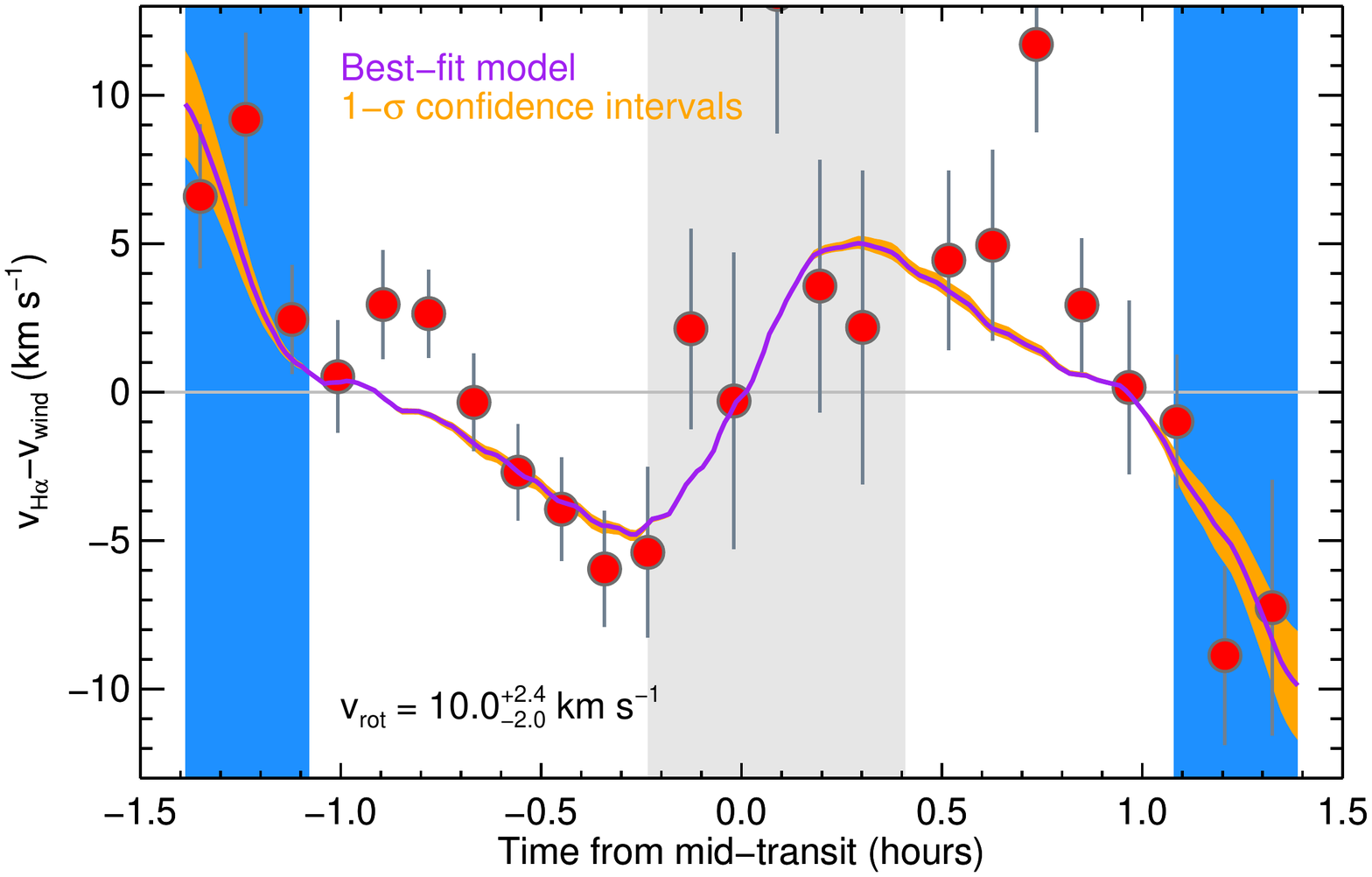}
   \figcaption{Same as \autoref{fig:havels} but now with the best-fit day-to-night
   side wind velocity $v_\text{wind}$ from \autoref{tab:mfits} removed. The 
   rotational model with the minimum $\chi^2$ value is shown in purple and the 
   estimated $1\sigma$ confidence intervals are shaded orange. Only the first and 
   last four velocities are included in the model comparison. The best-fit rotational 
   velocity of 8.5 km s$^{-1}$ is consistent with that found from the average 
   transmission spectrum fits.
\label{fig:velfits}}

\end{figure*}

To accomplish this we computed a grid of $v_{H\alpha}$ time series models using the
framework from \autoref{sec:mods}. We generated the models using the best-fit atmospheric
parameters from the TRW model in \autoref{sec:fits} with the
exception of letting $v_\text{rot}$ vary. We created models for 41 different
values of $v_\text{rot}$ ranging from 0.0 km s$^{-1}$ up to 20.0 km s$^{-1}$ in 
steps of $0.5$ km s$^{-1}$. We note that the velocity centroids are
mostly independent of the other model parameters besides $v_\text{rot}$ so using our best-fit
parameters from the average in-transit spectrum fit does not bias the
rotational velocities measured from the $v_{H\alpha}$ grid.

Since the transmission spectra contain the maximal amount of information
about atmospheric rotation upon ingress and egress, we only include the first
and last four $v_{H\alpha}$ measurements from \autoref{fig:havels} in the model
comparison. The in-transit points are equally consistent with all of the rotational
velocities since the line profile centroids at these times are dominated by
other effects. Thus the model grid comparison consists of $n=8$ data points
and $m=1$ free parameter ($v_\text{rot}$). Before comparing the data to
the model grid we subtract the best-fit value of $v_\text{wind}$ for the
TRW model ($v_\text{wind} = -4.6$ km s$^{-1}$) from the
measured $v_{H\alpha}$ values \autoref{tab:mfits}. In principle this leaves
only velocities contributions from the rotating atmosphere. 

Given the small number of data points and their large individual uncertainties, 
we opt for a simple $\chi^2$ comparison. For each $v_\text{rot}$ model we calculate
$\chi^2$ between the data and the model interpolated onto the mid-exposure
times of the eight selected points. The 68\% confidence interval on $v_\text{rot}$
is then calculated by taking all models for which $\chi^2 < \chi^2_\text{min} + 
\Delta\chi^2$ where $\Delta\chi^2 = 1.0$ in this case \citep{avni76}.

We plot the $v_{H\alpha}$ time series with the wind velocity removed in
\autoref{fig:velfits}, along with the best-fit rotational model (purple
line) and the $1\sigma$ confidence intervals. We have omitted the velocities
measured from the transmission spectra which overlap heavily with the
stellar RM profile (gray shaded region). The rotational model with $v_\text{rot} = 10.0^{+2.4}_{-2.0}$
km s$^{-1}$ provides the best fit to the data and traces the shape of the
$v_{H\alpha}$ measurements during ingress and egress. In addition, the evidence
for a rotating atmosphere is strengthened by the fact that the best-fit 
rotational velocity from the $v_{H\alpha}$ time series is consistent
at the $1\sigma$ level with the best-fit rotational velocity,
$v_\text{rot} = 10.1^{+0.8}_{-1.0}$ km s$^{-1}$, derived from the 
transmission spectrum fits.  

\section{Discussion}
\label{sec:disc}

We have demonstrated that the velocity centroids of the individual WASP-33 b \halpha\ 
transmission spectra show a pattern consistent with what is expected from
a rotating atmosphere. The rotational velocity of $v_\text{rot} = 10.1^{+0.8}_{-1.0}$ 
km s$^{-1}$ derived from the average transmission spectrum fit is larger than WASP-33 b's
tidally locked value of $v_\text{tl} = 7.1$ km s$^{-1}$. However, the velocity 
centroid time series analysis results in a smaller rotational velocity of 
$v_\text{rot} = 10.0^{+2.4}_{-2.0}$ which is consistent at the $2\sigma$ level with the planet's atmosphere 
rotating at the tidally locked rate. We also find that a large day-to-night side
wind velocity of $v_\text{wind} = -4.6^{+0.4}_{-0.4}$ km s$^{-1}$ is required
to explain the blue-shifted transmission spectra. The wind velocity becomes
less significant, $v_\text{wind} = -4.6^{+3.4}_{-3.4}$ km s$^{-1}$, when we consider 
uncertainties in the mid-transit time and how this affects the global 
velocity offset of the transmission spectra.

Measuring a planet's rotational velocity via its transmission spectrum is 
dependent on the assumption that the rotational axis is perpendicular to
the orbital plane, i.e., the planet has zero obliquity. Thus we are 
in fact constraining the value $v_\text{rot}$sin$\theta$cos$\phi$
where $\theta$ is the angle between the rotation axis and the planet's orbital
plane and $\phi$ is the angle between the rotation axis, projected onto
the orbital plane, and the line between the observer and the star. 
If $\theta = 0^\circ$ and $\phi=0^\circ$
then the ingress and egress velocity centroids will be $\approx 0$ km s$^{-1}$
since the atmosphere has no line-of-sight velocity relative to the observer. 
The transmission spectrum in this case will also exhibit no rotational broadening,
i.e., we are observing the planet pole-on. If $\theta = 90^\circ$ then the 
centroid velocities will be maximized during ingress and egress and the
average transmission spectrum broadening is maximized. An interesting intermediate
case is if $\theta = 0^\circ$ but $\phi = \pm 90^\circ$: the ingress and egress
velocities will zero since the planetary poles enter/exit the stellar disk 
first/last but the transmission spectrum broadening is still maximized
because the rotation axis is perpendicular to the line-of-sight.

Comparing the velocities derived from the two methods provides another possible avenue
for constraining the planet's obliquity: if the rotational broadening from the
average transmission spectrum is consistent with the time-series rotational
velocity, then $\theta \approx 90^\circ$. Although it's possible to
have consistent rotational velocities if $\theta$ has a value other than 90$^\circ$,
this would require $\phi \approx 0$ since any moderate value of $\phi$ would
decrease the ingress and egress velocities and increase the line
broadening. Additionally, we are far less likely to observe any particular 
planet at $\phi \approx 0$ if $\theta$ is small since most viewing angles,
given a fixed direction of the planet's spin vector, result in $\phi \neq 0.$ 

The only current empirical constraint
on the obliquity of a planetary mass companion was recently presented by
\citet{bryan20} who showed that the $12-27$ M$_\text{J}$ companion to 2MASS
J01225093-2439505 likely has a non-zero obliquity. Although we cannot
calculate a precise obliquity estimate for WASP-33 b, the agreement between 
the rotational velocities from our average spectrum model and the time-series 
grid model supports the conclusion that WASP-33 b's rotation axis is roughly perpendicular 
to its orbital plane, i.e., its obliquity is close to zero. This is
not unexpected since it is difficult for hot Jupiters to maintain
non-zero obliquity without the presence of perturbing companion \citep{millholland19}.
Even so, our analysis provides a rough constraint on the obliquity of a Jupiter-mass 
planet.

A second notable result from our analysis is the large blue-shift seen in
the transmission spectrum. Our model finds that a day-to-night side wind
of $v_\text{wind} = -4.6^{+3.4}_{-3.4}$ km s$^{-1}$ can account for this
shift. A large in-transit blue-shift of $-11 \pm 0.7$ km s$^{-1}$ was recently measured
for the UHJ WASP-76 b in the cross-correlation function of the planet's
neutral iron absorption lines \citep{ehrenreich20}. While the uncertainties
in our measured mid-transit timing prevent us from establishing the
absolute blue-shift of the transmission spectrum with more precision,
wind velocities of $\approx 5$ km s$^{-1}$ are not commonly seen for 
hot and ultra-hot planets. For example, \citet{yan18}, \citet{cauley19}, and \citet{wyttenbach20} 
all found no evidence of a velocity shift in the transmission spectrum of KELT-9 b,
currently the hottest transiting gas giant to be studied in detail. Most
measured transmission spectrum velocities in hot planet atmospheres are
on the order of $\approx 2$ km s$^{-1}$ \citep[e.g.,][]{wyttenbach15,casasayas17}.
Our results, combined with those from \citet{ehrenreich20}, hint that
the average day-to-night side wind speed increases into the ultra-hot
atmosphere regime \citep[$T_\text{eq} > 2000$;][]{parmentier18,baxter20}
but these flows are disrupted in the extreme limits of an atmosphere
like KELT-9 b's at $T_\text{eq} = 4000$ K. Indeed, \citet{komacek20}
showed that wind speeds on the order of $\approx 5-6$ km s$^{-1}$ are
possible at pressures of 1 mbar for UHJ atmospheres.

In their analysis of \ion{Ca}{2} H and K and infrared triplet absorption
in WASP-33 b's atmosphere \citet{yan19} do not see a consistent blue-shift
in the two sets of lines. A blue-shift similar in magnitude to what we measure 
would likely have been detected in their investigation. The same 
caveats with regards to the mid-transit time uncertainty apply to their procedure
so it is possible that the true blue-shift is simply masked by
inherent uncertainties in the adopted mid-transit time.
The Balmer lines and the \ion{Ca}{2} lines are expected to form
at similar pressures in UHJ atmospheres so we do not expect the 
velocity flows sampled by the lines to differ significantly \citep{turner20}. Given the
discrepancies between our velocity measurement and those from \citet{yan19}
we suggest that a concrete interpretation of the day-to-night side wind
signature await additional spectroscopic transit studies. High-quality
simultaneous photometry would aid in reducing the mid-transit time
uncertainty for the spectroscopic analysis.

Although our models are able to account for the morphology and velocity
centroids of WASP-33 b's Balmer line transmission spectra, a more realistic
treatment of velocity flows in the planet's thermosphere could reveal
important information about which velocity features contribute to the line
profiles \citep[e.g.,][]{flowers19}. For example, the atmosphere is likely 
not rigidly rotating from the stratosphere out to the edge of the thermosphere 
and jets, which are a ubiquitous feature of hot planet GCMs
\citep[e.g.,][]{rauscher10,showman13,carone20,komacek20},
probably contribute to the transmission spectrum even though they are
not discernible in our model treatment of the line profiles. 

Finally, we emphasize the need for GCM exploration
of the uppermost bound atmospheric layers of hot and ultra-hot planets.
Such simulations are currently difficult to achieve primarily due to the
breakdown of the primitive equations of meteorology at micro- and nano-bar
pressures. Atomic absorption lines in hot planet atmospheres generally trace these
low-pressure layers and they are relatively straightforward to measure.
Thus the rapidly growing number of hot planet atmosphere detections, 
and their associated velocity signatures, needs
a better theoretical platform upon which to interpret any observed velocity
signatures rather than extrapolating from the abundance of GCM results
at higher pressures. 

\section{Conclusions}
\label{sec:conclusions}

We have presented the first detections of the Balmer lines \halpha\ 
and \hbeta\ in the atmosphere of WASP-33 b and measured the velocity
centroids of the individual \halpha\ transmission spectra as a function
of time throughout the transit. Both the velocity centroids 
($v_\text{rot} = 10.0^{+2.4}_{-2.0}$ km s$^{-1}$) and the
rotational broadening measured from the in-transit spectrum 
($v_\text{rot} = 10.1^{+0.8}_{-1.0}$ km s$^{-1}$) are 
consistent with a super-rotating atmosphere, although the time-series 
\halpha\ velocities only differ by $2\sigma$ from the expected tidally-locked 
rotation rate of WASP-33 b. We do not find any 
evidence of velocity signatures associated with equatorial jets, although 
such features may be swamped by the planet's fast rotation and possibly not
retrievable using the Balmer line transmission spectra. 

One important consideration to reiterate about our results is
the potential bias introduced by stellar pulsations. We have attempted to
simulate the effect of the pulsations on the transmission spectrum
line profile and the measured velocity centroids. We find that pulses
can have a moderate effect on the retrieved model parameters and, on
average, shift the centroid velocities by $\approx 2.2$ km s$^{-1}$.
Thus while we don't expect the results of our analysis to be strongly
biased by pulsation effects, their presence should be taken into account
when adopting particular model values derived from our work. One possible
way to address the pulsation effects is multiple transits: the pulsation
signal should overlap with different pieces of the planetary absorption
from transit to transit. This could help further clarify how strongly
the pulses affect the transmission spectrum.

Our observations reinforce the power of time-resolved transmission
spectra where the signal-to-noise of individual exposures permit reliable
extraction of the profile morphologies and velocity centroids. Similar
measurements can be performed with the cross-correlation technique when
many transitions of an element are present in the planet's atmosphere but
each one is too weak to analyze independently \citep[e.g.,][]{ehrenreich20,borsa20}. Time-series
measurements of velocity features in hot planet atmospheres can help
resolve the line broadening degeneracy between, for example, spherical 
expansion and rotation due to the very different velocity time-series
produced by each feature. The advent of 30-meter telescopes will greatly
expand the parameter space for which these experiments are possible
and may enable variability studies \citep[e.g.,][]{komacek20} of velocity
flows in the atmospheres of transiting exoplanets. Empirical constraints
on atmospheric velocity dynamics are a critical part of our ongoing
pursuit of a more complete understanding of the physics of exoplanets
and should be extracted when possible. 

\acknowledgments

{We thank the referee for a thorough review of the manuscript and their helpful suggestions,
which significantly improved the quality of the work. P.W.C. is grateful to Sarah 
Millholland for a conversation about the obliquity evolution of hot Jupiters. This research 
has made use of the NASA Exoplanet Archive, which is operated by the
California Institute of Technology, under contract with the National Aeronautics and
Space Administration under the Exoplanet Exploration Program. Additionally, this work
has made use of NASA's Astrophysical Data System and of the SIMBAD database, which is
operated at CDS, Strasbourg, France. This work has also made use of the VALD database,
operated at Uppsala University, the Institute of Astronomy RAS in Moscow, and the
University of Vienna. We thank the staff at the Large Binocular Telescope for their 
role in collecting the data presented in this manuscript.}

\software{\texttt{EXOFAST}, \citet{eastman13}; \texttt{Molecfit}, \citet{kausch15}; \texttt{Spectroscopy Made Easy
(SME)}, \citet{valenti96,piskunov17}}

\bibliography{references}{}
\bibliographystyle{aasjournal}

\appendix
\label{app:appendix}

\section{Individual transmission spectra plots}
In this appendix we provide the individual in-transit transmission spectra for
\halpha\ and \hbeta in the rest frame of the planet. All \halpha\ spectra have been binned 
by a factor of 2 and the \hbeta\ spectra were binned by a factor of 5 for clarity.

\begin{figure*}[htbp]
   \centering
   \includegraphics[scale=.9,clip,trim=20mm 70mm 0mm 25mm,angle=0]{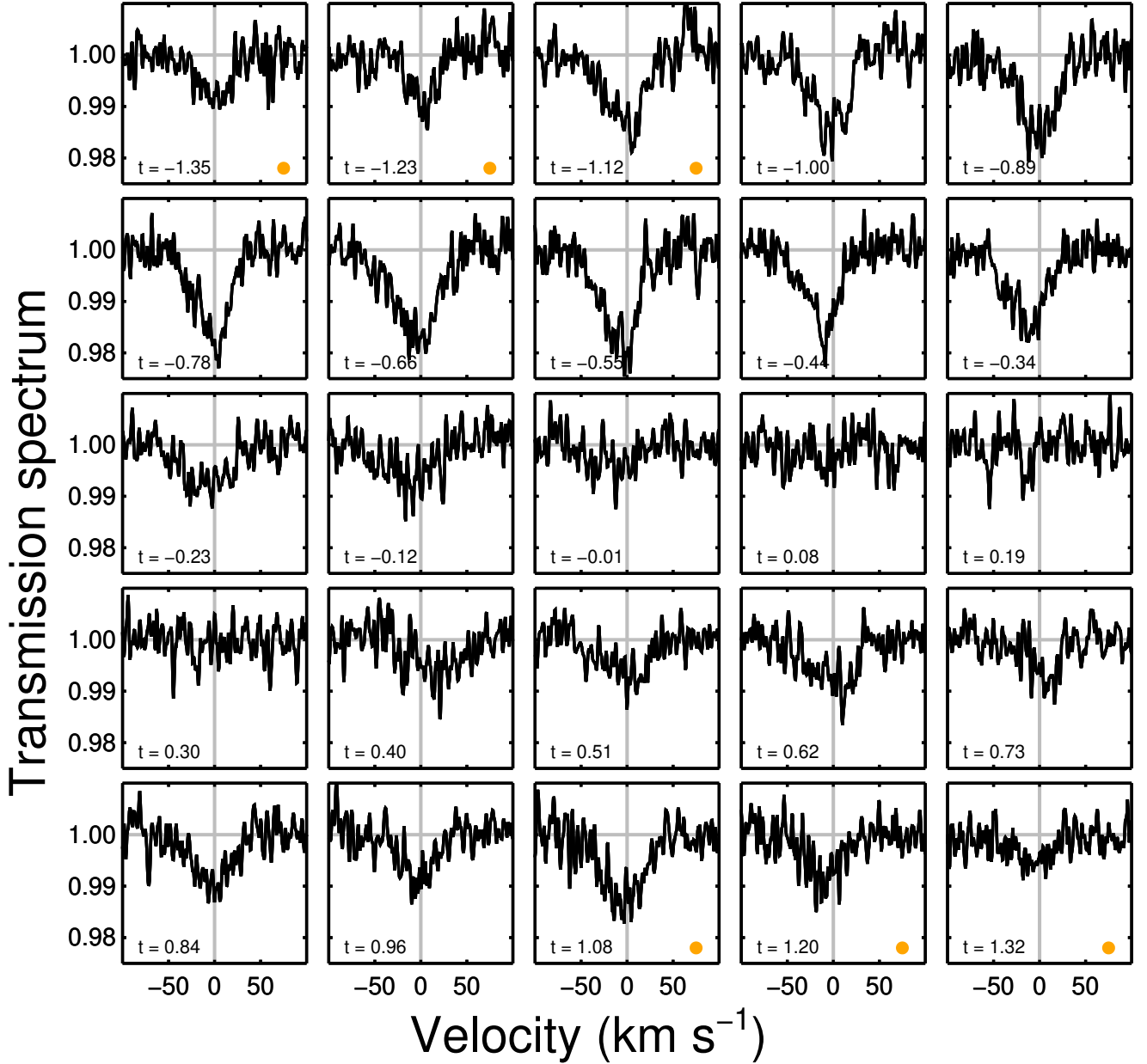}
   \figcaption{In-transit H$\alpha$ transmission spectra in the planetary rest frame. 
   Mid-exposure times are marked in hours in the bottom left of each panel. The vertical 
   gray line marks $v = 0.0$ km s$^{-1}$. Exposures taken during ingress and egress are identified
   with an orange circle in the lower right of each panel. Note that the spectra are
   not corrected for the velocity offset identified in \autoref{sec:modresults}.
\label{fig:tspecsha}}

\end{figure*}

\begin{figure*}[htbp]
   \centering
   \includegraphics[scale=.9,clip,trim=20mm 70mm 0mm 25mm,angle=0]{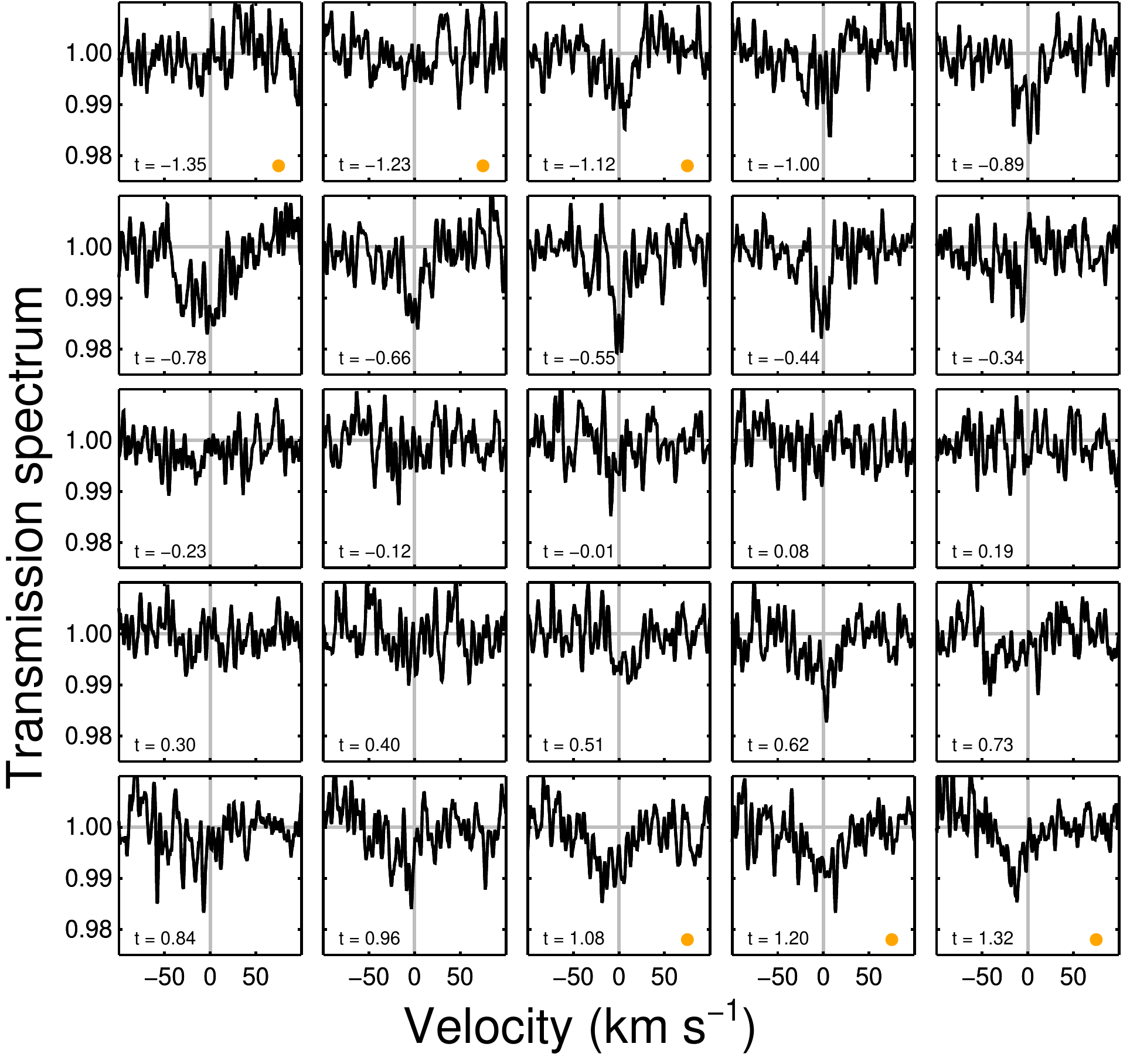}
   \figcaption{Same as \autoref{fig:tspecsha} but for H$\beta$. 
\label{fig:tspecshb}}

\end{figure*}

\end{document}